\documentclass[3p, times]{elsarticle}
\title{A semi-analytical geometrical acoustics method for numerical simulation of ultrasound based motion sensing}
\author{Vamshi Krishna Chillara \corref{cor1}}
\ead{vamchill@amazon.com}
\author{Wontak Kim}
\cortext[cor1]{Corresponding author: Vamshi Krishna Chillara (vamchill@amazon.com)}
\address{Acoustics Reseach and Algorithms Team, Edge Technology Group, Amazon}
\address{101 Main Street, Cambridge, MA, 02142, USA}
\usepackage{amsmath}
\usepackage{color}
\usepackage{subcaption}
\usepackage{float}
\usepackage{hyperref}
\graphicspath{{./Figures_publish/}}
\begin{document}
	
\begin{abstract}
	We present a semi-analytical geometrical acoustics method to numerically simulate ultrasonic signal characteristics pertinent to motion sensing applications in indoor environments. The proposed methodology treats motion sensing from the first-principles in the sense that the expressions for acoustic field from the source, that scattered by the target and then received at the receiver are all derived from a kinematic standpoint incorporarting target motion into consideration. Thus, the method outputs the time-domain acoustic pressure field at the receiver whose analytic signal representation has both amplitude (AM) and phase/frequency (FM) modulation. Doppler effect is then derived to be an outcome of computing the instantaneous frequency of the resultant analytic signal.  In this sense, it differs from some of the existing methods for modeling the Doppler effect where the Doppler frequency shifts are first computed and then artificially introduced in postprocessing phase by either time/frequency scaling of the received signal without recourse to presence of multiple reflections or the finite extent of targets resulting in a frequency spread of Doppler shift in realistic scenarios.  In addition to the frequency shifts observed from the Doppler effect, we show that our methodology accurately captures the signal strength variations for cases with non-uniform source directivity, presence of multiple reflections, and with different target motion velocity profiles/trajectories. We first present the method for simple cases, namely 1) moving receiver in the pressure field of a stationary source, 2) stationary receiver in the pressure field of a moving source, and 3) a point scatter moving in the presence of stationary source and receiver. For each of these cases, explicit  expressions for ultrasonic Doppler signal are presented with the corresponding analytic signal representations giving amplitude and phase/frequency modulation functions. The Doppler effect is explicitly demonstrated by computing the instantaneous frequency of the signal for a few simple cases. The dervied expressions from the proposed methodology agree with the ones available in literature for these simple cases. Then, we extend the methodology for i) sources with non-uniform directivities ii) environments with multiple reflections and iii) incorporate extended targets as point-clouds with orientation dependent scattering response functions. We briefly discuss the  Image Source Method (ISM)  and integrate it with the above methodology to demonstrate the Doppler signal characteristics in  environments with reflections. The proposed methodology is numerically implemented in baseband frequency domain to facilitate simulation at  typical target motion time-scales (sec to mins) that are much longer than the ultrasonic frequency time scales ($\leq 0.05$ ms) where running the simulations using conventional methodologies are practically infeasible.   A series of examples are presented throughout to demonstrate the effect of  source directivity, wall reflections, and motion trajectories on the Doppler signal strength and frequency characteristics observed for motion sensing applications. Finally, we present a comparison of simulated results with experimental results on data acquired with a human target moving in an environment with an ultrasonic source and receiver. We specifically compare the baseband signal characteristics and their corresponding Short-time Fourier Transforms that depict Doppler frequency characteristics and show them to be in good qualitative agreement.

\end{abstract}	

\maketitle
\section{Introduction \label{sec1} } 
Motion sensing is ubiquitous and has several practical applications ranging from underwater acoustics\cite{wan2012performance}, indoor presence/occupancy detection\cite{hammoud2017ultrasense}, moving target ranging and localization\cite{rafa2008influence}, wearable sensing \cite{yang2020simultaneous, cheng2024usee} and gesture recognition \cite{ling2020ultragesture,watanabe2013ultrasound}. Both active and passive sensing technologies exist for these applications. Radio, Ultrasound, and Infrared that require emissions belong to active sensing category  and Computer Vision, ambient light sensors belong to the passive category.  Active sensing technologies rely on emissions from a source and anlayze reflections to detect target motion. Of these, radar and ultrasound are the most widely used technologies for motion sensing. While radar uses electromagnetic waves, of interest to this article is ultrasound based motion-sensing that utilize acoustic waves at ultrasonic frequencies ($\geq$20 kHz) that propagate in the medium and interact with objects in the environment. The reflected signals are then processed to decipher motion signatures. One of the most common motion signatures used for detection is the presence of the Doppler effect \cite{doppler1903ueber} in the received signal. It is one of the widely used technology for both radar \cite{tesch2015rfid} and ultrasound motion sensing \cite{sabatier2006ultrasonic, ekimov2008human}. This technology relies on the Doppler effect i.e., change in the apparent frequency of the signal at  receiver compared to the emission frequency from the source. In a 1D setting, where source and receiver are moving in a straight line the Doppler effect is described by the relation 
\begin{equation} \label{eqn0}
	\omega = \omega_c \left(\dfrac{c\pm v_m}{c\mp v_s}\right)
\end{equation}
where $\omega_c$ is the frequency of emission from the source, $v_m$ is the speed of the receiver, $v_s$ is the speed of the source and $c$ is the speed of the sound or the electromagnetic wave depending on the sensing mechanism used. For a general case of motion in 3D, the Doppler effect is described by more complex relations and has extensively been discussed in the literature \cite{klinaku2019doppler,alejos2023classical}. The focus of this paper is not on deriving accurate expressions for Doppler frequency but developing a numerical simulation methodology for ultrasound Doppler based motion sensing.

Practical applications that use ultrasound for Doppler-based motion sensing emit an ultrasonic carrier wave from the source and receive the reflected signal from the environment at the receiver. The Doppler effect is inferred by processing the received signal to discover inherent attributes pertaining to the motion and environment.  Our overarching goal is to develop a numerical simulation methodology for ultrasound Doppler based motion sensing for indoor room-acoustic applications.  This goal stems from the need to develop advanced signal processing and machine learning techniques that rely on large quantities of data to train models under varying realistic conditions such as in the presence of external noises, complex motion characteristics and different source/receiver type and positions. Numerical simulation offers a reliable way to generate such large quantities of data eliminating the need for manual data collection which incurs significant time/resource costs. Developing such a simulation methodology requires:
\begin{enumerate}
	\item Incorporating realistic acoustic sources that have a non-uniform spatial directivity as opposed to point sources that have uniform directivity. Non-uniform directivity results in field variations that are very apparent in the signal received at the receiver and are directly relevant to Signal-to-Noise ratio (SNR) for sensing applications.
	\item Incorporating multiple reflections in the source/target/receiver environment. Doppler signal characteristics for multi-reflection environements can be very different than for the cases where there is just a single direct-path signal from the source to the target.
	\item Running simulations at high source emission frequencies ($\geq$ 20 kHz) that may require high sampling.
	\item Integrating complex target motions and trajectories into the simulation framework.
\end{enumerate}
General time-domain methods are challenging to be implemented for room-acoustic applications at ultrasonic frequencies due to large domain sizes ($\approx$ 5 m) relative to the wavelength of the ultrasonic wave ($\approx$8 mm at 40 kHz). In addition, incorporating target motion into a simulation is challenging as they can typically have time-varying speeds/motion trajectories and more importantly longer time-scales (sec to mins) than those for sampling ultrasonic waves ($\leq 0.05$ ms). So, Geometrical Acoustics (GA) is a more practical choice for these simulations. In this article, we use the Image Source Method from room-acoustics and combine with the semi-analytical formulation for simulating the Doppler signal developed in the article to demonstrate an end-to-end simulation framework for Ultrasonic Doppler based motion sensing applications. To our knowledge, such methodology has not been investigated or presented in the literature for motion sensing applications. The key contributions of this article are:
\begin{enumerate}
	\item Semi-analytical method for computing the received ultrasonic signal from moving targets that have spatial extent (i.e., not point scatterers/targets).
	\item A methodology to handle multiple reflections by integrating the proposed method with the Image Source method. 
	\item Ability to incorporate different target motion velocity profiles and trajectories into the simulation framework.
	\item A compute-efficient way to run the simulations in the baseband signal domain that is amenable to running large number of simulations and with longer target motion duration.
\end{enumerate}

The content of the article is organized as follows. We first present the semi-analytical formulation for computing the received signal for moving point receviers and point sources in sections \ref{sec2} and \ref{sec3} respectively. Then we develop analytical expressions for a point scatterer moving in an environment with a stationary source and receiver in section \ref{sec4}. In section \ref{sec5}, we discuss how we incorporate the source directivity and multiple reflections into this framework using the image source method. We also highlight the advantage of implementing  the methodology in baseband domain to reduce the computation costs associated with temporal sampling of ultrasonic signals. In section \ref{sec6}, we present numerical simulation results for moving targets in a room environment with reflections. We first discuss some key aspects of the simulation, namely the influence of velocity profiles, influence of reflections and  motion trajectories on Doppler signal characteristics using examples from the motion of a point target in the room environment. Then, we discuss a methodology to incorporate finite extended targets into the simulation framework by representing them as point clouds with orientation dependent scattering response functions.  We present a few sample results to illustrate the differences from modeling a point source versus an extended target. In section \ref{sec7}, we compare the results from simulation with experiments where the Doppler signal characteristics are shown to be in good qualitative agreement.  Finally, in section \ref{sec8}, we present the conclusions and discuss how this approach can be extended and improved to incorporate more realistic target motion representations.

\section{Receiver moving in the pressure field of a stationary source \label{sec2} }  The mathematical notation used in this article is shown in Table \ref{tab5}.
\begin{table}[H]
	\begin{center}
		\caption{Notation used in the article \label{tab5}}
			\begin{tabular} {c c}
					\hline\hline
					$\vec{r}_s(t), \vec{r}_s$ & Position of the source \\
					$\vec{r}_m(t), \vec{r}_m$ & Position of the  receiver\\
					$\vec{r}_p(t)$ & Position of the moving target\\
					$\vec{v}_p(t)$ & Velocity of the moving target\\
					$\vec{k}$ & Wavevector\\
					$\omega_c$ & Carrier frequency\\
					$p(\vec{r},t)$ & Time-dependent pressure field \\
					${s}_{m}(t)$ & Signal at the receiver
				\end{tabular}
		\end{center}
\end{table}

In this section, we consider the case of a moving receiver in the pressure field of a stationary source. We assume the pressure field generated by the source is given by $p(\vec{r},t)$. Also, we assume that the receiver response is instantaneous to changes in the pressure field and the motion of the receiver does not cause significant changes to the pressure field. Under these assumptions, if the motion of the receiver is prescribed by $\vec{r}_m(t)$, then the time-domain signal received at the receiver is given by
\begin{equation}\label{eqn1}
	{s}_{m}(t) = p(\vec{r}_m(t),t)
\end{equation}
Note that ${s}_{m}(t)$ is the time-domain signal as recorded by the receiver and will include both the carrier and Doppler-shifted components of the signal. We consider the analytic signal representation of the ${s}_{m}(t)$ given by 
\begin{equation} \label{eqn2}
	{s}_{m}(t) = A_{m}(t) e^{i\phi_m(t)}
\end{equation}
where $A_m(t)$ is the instantaneous amplitude and $\phi_m(t)$ is the instantaneous phase of the analytic signal. The instantaneous angular frequency of the signal, ${s}_{m}(t)$, can be obtained as
\begin{equation}\label{eqn3}
	\omega(t)= \frac{d\phi_m(t)}{dt}
\end{equation}

We now consider some simple cases of the pressure field, $p(\vec{r},t)$,  to demonstrate how the Doppler effect manisfests in the received signal of a moving receiver. For brevity, all the pressure fields in this article are represented as complex exponentials and these quantities should be interpretted as the real part, $Re\{ .\}$ of the complex-valued representation.

\subsection{Receiver moving in a planewave field}
Consider a planewave propagating in the medium with a pressure field given by 
\begin{equation} \label{eqn4}
	p(\vec{r},t) = e^{i\left( \omega_c t - \vec{k}_{0}.\vec{r}\right)}
\end{equation}
where $\omega_c$ is the carrier frequency and $\vec{k}_{0}$ is the wave-vector defining the direction of planewave propagation. In the pressure field of the planewave, consider a point receiver moving with a motion trajectory defined by $\vec{r}_m(t)$. The signal at the receiver for this case (from Eqn(\ref{eqn1})) is given by
\begin{equation}\label{eqn5}
	{s}_{m}(t) = p(\vec{r}_m(t),t) = e^{i\left( \omega_c t - \vec{k}_{0}.\vec{r}_{m}(t)\right)}
\end{equation}
Comparing Eqn(\ref{eqn5}) with Eqn(\ref{eqn2}), we have $A_m(t) = 1$ and $\phi_m(t) = \left(\omega_c t - \vec{k}_{0}.\vec{r}_{m}(t)\right)$ and thus

\begin{equation}\label{eqn6}
	\omega(t) = \frac{d\phi_m(t)}{dt} = \left(\omega_c -\vec{k}_{0}.\frac{d\vec{r}_{m}(t)}{dt} \right)=\left(\omega_c -\vec{k}_{0}.\vec{v}_{m}(t) \right)
\end{equation}
where $\vec{v}_{m}(t) = \dfrac{d\vec{r}_{m}(t)}{dt}$ is the velocity of the moving receiver.

From the above equation, the frequency as seen by the receiver is dependent on its instantaneous motion relative to the wavevector, $\vec{k}_{0}$ , of the planewave.
For special cases where the receiver is moving in the same direction of the wave, moving perpendicular to the direction of the planewave, and moving opposite to the direction of propogation of the planewave, the instantaneous frequncies are given by 
\begin{equation}\label{eqn7}
	\omega  =  \begin{cases} 
		\omega_c - ||\vec{k}_0||||\vec{v}_m|| & \text{if } \vec{v}_{m} \hspace{4pt} \text{and} \hspace{4pt} \vec{k}_{0} \hspace{4pt} \text{are parallel i.e., receiver moving in the same direction as the planewave}  \\
		\omega_c  & \text{if } \vec{v}_{m} \hspace{4pt} \text{and} \hspace{4pt} \vec{k}_{0} \hspace{4pt} \text{are orthogonal i.e., receiver moving orthogonal to the planewave}\\
		\omega_c + ||\vec{k}_0||||\vec{v}_m|| & \text{if } \vec{v}_{m} \hspace{4pt} \text{and} \hspace{4pt} \vec{k}_{0} \hspace{4pt} \text{are antiparallel i.e., receiver moving towards the planewave}
	\end{cases}
\end{equation}

Note that the above expressions are identical to the ones typically presented for Doppler effect in 1D.

\subsection{Receiver moving in the pressure field of an isotropic point source}
Consider a point source with uniform directivity and strength, $p_0$, at the location, $\vec{r}_s$, radiating pressure field with  frequency, $\omega_c$, in the medium. The pressure field generated by such a source is given by  
\begin{equation} \label{eqn8}
	p(\vec{r},t) = \frac{p_0}{||\vec{r}-\vec{r}_s||}e^{i\omega_c \left( t - \frac{||\vec{r}-\vec{r}_s||}{c}\right)}
\end{equation}
So, from Eqn(\ref{eqn1}), the signal received at the receiver for this case is given by 
\begin{equation} \label{eqn9}
	p(\vec{r}_m(t),t) = \frac{p_0}{||\vec{r}_m(t)-\vec{r}_s||}e^{i\omega_c \left( t - \frac{||\vec{r}_m(t) - \vec{r}_s||}{c}\right)}
\end{equation}
and so the analytic signal amplitude $A_m(t) =\frac{1}{||\vec{r}_m(t)-\vec{r}_s||}$ and the phase $ \phi_m(t) = \omega_c \left( t - \frac{||\vec{r}_m(t)-\vec{r}_s||}{c}\right)$.
The instantaneous frequency is thus
\begin{equation} \label{eqn10}
	\omega = \omega_c - \dfrac{\omega_c}{c}\dfrac{d||\vec{r}_m(t)-\vec{r}_s||}{dt}
\end{equation}
The Doppler shift, thus is given by 
\begin{equation} \label{eqn11}
	\omega - \omega_c = -{\omega_c}\dfrac{d}{dt} \left(\dfrac{||\vec{r}_m(t)-\vec{r}_s||}{c}\right)
\end{equation}
is thus proportional to the instantaneous rate of change of the time-of-flight of the sound, $\dfrac{d}{dt} \left(\dfrac{||\vec{r}_m(t)-\vec{r}_s||}{c}\right)$, from source to receiver. This is an important observation and will be useful to relate time-domain and frequency-domain methods for simulating motion sensing. In addition, this observation is the key to incorporating Doppler effects in audibile signals and has been used for realistic rendering of sounds for moving targets\cite{iwaya2007rendering} in acoustic environments.


\section{Stationary receiver in a pressure field generated by a moving point source \label{sec3}}
We consider a stationary receiver in a pressure field at the location $\vec{r}_{m}$. Consider a point source moving along a line with a constant velocity $\vec{v}_0$ starting at the origin at $t=0$ and pulsating at frequncy, $\omega_c$ . The position of the source as a function of time in this case is given by
\begin{equation}\label{eqn14}	
	\vec{r}_{s}(t) = \vec{v}_0 t
\end{equation}
The field generated by the point source is given by \cite{hongbin}
\begin{equation}\label{eqn15}	
	p(\vec{r}, t) =  \int_{t_0}^{t} \dfrac{p_0}{||\vec{r} - \vec{v}_0 \tau||} e^{i\omega_c \tau}  \delta\left(t- \dfrac{||\vec{r} - \vec{v}_0 \tau||} {c} - \tau \right) d\tau
\end{equation}
Here $\tau$ is called the retarted time. Physically, `$\tau$' represents the time instant  at which an emission emitted at the source reaches the point $\vec{r}$ at the time instant `$t$'. If the source were at rest then $\tau = t-\dfrac{||\vec{r}||}{c}$. In general, the relation between $\tau$ and $t$ depends on the kind of motion of the source and there is a need to solve an implicit equation as will be shown below.

In Eqn(\ref{eqn15}), if we let $g(\tau) = t- \dfrac{||\vec{r} - \vec{v}_0 \tau||} {c} - \tau$, then the above integral can be written as 
\begin{equation}\label{eqn16}	
	p(\vec{r}, t) =  \int_{0}^{t} \dfrac{p_0}{g'(\tau)||\vec{r} - \vec{v}_0 \tau||} e^{i\omega_c \tau}  \delta\left(g(\tau)\right) dg(\tau)
\end{equation}

This is equivalent to 
\begin{equation}\label{eqn17}	
	p(\vec{r}, t)  = \sum_{n}^{} \dfrac{p_0}{g'(\tau_n)||\vec{r} - \vec{v}_0 \tau_n||}e^{i\omega_c \tau_n} \text{such that}  \hspace{4pt} g(\tau_n)=0
\end{equation}

We can compute $g'(\tau)$ as
\begin{eqnarray}\label{eqn18}	
	g'(\tau) &=& \dfrac{(\vec{r}-\vec{v}_0\tau).\vec{v}_0}{c||\vec{r}-\vec{v}_0\tau||} -1 \\ \nonumber
	              &=& \dfrac{\vec{v}_0}{c}.\dfrac{\vec{r}-\vec{v}_0 \tau}{||\vec{r}-\vec{v}_0 \tau||} -1\\ \nonumber
	              &=& -\left(1-M_0 \cos(\theta_0(\tau))\right)
\end{eqnarray}
where $M_0 = \frac{||\vec{v}_0||}{c}$ is the Mach number and $\theta_0(\tau) = \dfrac{(\vec{r}-\vec{v}_0\tau)}{||\vec{r}-\vec{v}_0\tau||}.\dfrac{\vec{v}_0} {||\vec{v}_0||}$ is the angle between the moving source velocity vector and the instantaneous vector from the source to receiver location. 

For subsonic source speeds, $||\vec{v}_0|| < c$, there is only one $\tau$ for which $g(\tau)=0$. Also, if $||\vec{v}_0|| <<c$, then $t\approx\tau$ and we can approximate $g(\tau)$ using Taylor series expansion about $\tau = t$ as follows

\begin{eqnarray}\label{eqn19}	
	g(\tau)&=&g(t)+g'(t)\left(\tau-t\right)\\ \nonumber
	           &=&-\dfrac{||\vec{r} - \vec{v}_0 t||}{c} - \left(1-M_0 \cos(\theta_0(t)) \right)  \left(\tau-t\right)
\end{eqnarray}
where $M_0 = \frac{||\vec{v}_0||}{c}$ is the Mach number and $\cos(\theta_0(t)) = \dfrac{(\vec{r}-\vec{v}_0t)}{||\vec{r}-\vec{v}_0t||}.\dfrac{\vec{v}_0} {||\vec{v}_0||}$.

From Eqn(\ref{eqn19}), solving for $\tau$ yields
\begin{equation}\label{eqn20}	
	\tau= t -\dfrac{1}{1-M_0 \cos(\theta_0(t)) }\dfrac{||\vec{r} - \vec{v}_0 t||}{c} 
\end{equation}
Eqn(\ref{eqn20}) gives explicitly the retarted/emission time as a function of $t$.  Note that $\tau < t$ indicating causality i.e., emission time is always before the receiving time for all cases. More accurate expressions for $\tau$ can be derived by considering the second order approximation for the Taylor series used in Eqn(\ref{eqn20}) but the above expressions suffice for the source/target speeds encountered for motion sensing applications considered in this article.

Using Eqns(\ref{eqn17} \& \ref{eqn20}), we get the field at the point $\vec{r}$ at time $t$ is given by 

\begin{equation}\label{eqn21}	
	p(\vec{r}, t)  = \dfrac{1}{1-M_0 \cos\left(\theta_0(t)\right)} \dfrac{p_0}{\left(||\vec{r} - \vec{v}_0 t||\right)} e^{i\omega_c \left( t -\dfrac{1}{1-M_0 \cos(\theta_0(t)) }\dfrac{||\vec{r} - \vec{v}_0 t||}{c} \right)}
\end{equation}
Note that while the above expression is dervied for constant velocity, it is equally valid for a general time-varying velocity profile provided ${||\vec{v}_0||}<<c$ and one can replace $M_0$ by the corresponding speed at that time instant i.e., $M_0= \frac{||\vec{v}_0(t)||}{c}$. For the stationary receiver at $\vec{r}_m$, the field recorded by it will be given by 
\begin{equation}\label{eqn22}	
	p(\vec{r}, t)  = \dfrac{1}{1-M_0 \cos\left(\theta_0(t)\right)} \dfrac{p_0}{\left(||\vec{r}_m - \vec{r}_s (t)||\right)} e^{i\omega_c \left( t -\dfrac{1}{1-M_0 \cos(\theta_0(t)) }\dfrac{||\vec{r}_m - \vec{r}_s (t)||}{c} \right)}
\end{equation}
where we replaced $\vec{v}_0t=\vec{r}_s(t)$, the instantaneous position of the source.

We next look at the anlaytic signal representation of the above equation, Eqn(\ref{eqn22}), to obtain

\begin{equation}\label{eqn23}
	A_m(t)=\dfrac{1}{1-M_0 \cos\left(\theta_0(t)\right)} \dfrac{p_0}{\left(||\vec{r}_m - \vec{r}_s (t)||\right)}
\end{equation}

\begin{equation}\label{eqn24}
	\phi_m(t)=\omega_c \left( t -\dfrac{1}{1-M_0 \cos(\theta_0(t)) }\dfrac{||\vec{r}_m - \vec{r}_s (t)||}{c} \right)
\end{equation}
\begin{equation}\label{eqn25}
	\omega(t)= \dfrac{d\phi(t)}{dt} \approx \omega_c -\dfrac{1}{1-M_0 \cos(\theta_0(t))} \dfrac{d}{dt} \left(\dfrac{||\vec{r}_m - \vec{r}_s (t)||}{c}\right)
\end{equation}
The above expressions are the general expressions valid for any source motion with the receiver at rest under the approximation of $||\vec{v}_0(t)||<<c$. It should be noted the source motion will result in both instantaneous amplitude and phase variations. Compared to the case where the source is stationary and receiver is in motion, the amplitude for the source motion has an additional factor,  $\dfrac{1}{1-M_0 \cos(\theta_0(t))}$. Compared to Eqn(\ref{eqn11}) for the receiver in motion, here  the instantaneous frequency, $\omega(t)$, is not just dependent on the rate of change of instantaneous time-of-flight but is also multiplied by the factor, $\dfrac{1}{1-M_0 \cos(\theta_0(t))}$, highlighting the difference between source and receiver motions for the Doppler shift. We now present a few special cases where these expressions reduce to well-known expressions for Doppler effect.
\subsection{Doppler effect for a few simple cases}

\begin{itemize}
	\item Source and receiver stationary: For this case, we have $M_0=0$. From Eqn(\ref{eqn23} \& \ref{eqn24}), we get 
	\begin{equation}\label{eqn26}
		A_m(t)= \dfrac{p_0}{\left(||\vec{r}_m - \vec{r}_s ||\right)}
	\end{equation}
	\begin{equation}\label{eqn27}
		\phi_m(t)=\omega_c \left( t - \dfrac{||\vec{r}_m - \vec{r}_s||}{c} \right)
	\end{equation}
	\begin{equation}
		\omega(t)= \dfrac{d\phi(t)}{dt} = \omega_c
	\end{equation}
	
	\item Source moving towards the stationary receiver along a straight line: For this case, we have $\theta_0=0^{\circ}$. From Eqn(\ref{eqn23} \& \ref{eqn24}), we get 
	\begin{equation}\label{eqn28}
		A_m(t)= \dfrac{1}{1-M_0}  \dfrac{p_0}{\left(||\vec{r}_m - \vec{r}_s(t) ||\right)}
	\end{equation}
	\begin{equation}\label{eqn29}
		\phi_m(t)=\omega_c \left( t - \dfrac{1}{1-M_0} \dfrac{||\vec{r}_m - \vec{r}_s(t)||}{c} \right)
	\end{equation}
	
	\begin{equation} \label{eqn30}
		\omega(t)= \dfrac{d\phi(t)}{dt} = \omega_c -\omega_c \dfrac{-M_0}{1-M_0}  =\dfrac{\omega_c}{1-M_0}=\omega_c \dfrac{c}{c-||\vec{v}_0||}
	\end{equation}
	\item Source moving away from the stationary receiver:
	$\theta_0=180^{\circ}$. From Eqn(\ref{eqn23} \& \ref{eqn24}), we get 
	\begin{equation}\label{eqn31}
		A_m(t)= \dfrac{1}{1+M_0}  \dfrac{p_0}{\left(||\vec{r}_m - \vec{r}_s(t) ||\right)}
	\end{equation}
	\begin{equation}\label{eqn32}
		\phi_m(t)=\omega_c \left( t - \dfrac{1}{1+M_0} \dfrac{||\vec{r}_m - \vec{r}_s(t)||}{c} \right)
	\end{equation}
	
	\begin{equation} \label{eqn33}
		\omega(t)= \dfrac{d\phi(t)}{dt} = \omega_c -\omega_c \dfrac{M_0}{1+M_0}  =\dfrac{\omega_c}{1+M_0}=\omega_c \dfrac{c}{c+||\vec{v}_0||}
	\end{equation}
\end{itemize}

The above expressions for each of the cases correspond to the standard expressions for Doppler effect that are commonly encountered for the source motion.

\section{Point scatterer moving in the pressure field of a source and reflecting ultrasound towards receiver \label{sec4}}
Consider  a point scatterer moving in the pressure field generated by a stationary source pulsating with frequency, $\omega_c$, with a stationary point receiver in the medium. The medium is assumed to be free from any reflectors other than the scatterer. Both the source and receiver are stationary. We are interested in the time-dependent pressure signal, $s_m(t)$, at the receiver as the scatterer moves in the domain. We make the following observations in this regard:
\begin{enumerate}
	\item In the absence of the scatterer, the receiver would receive the pressure field at the same carrier frequency $\omega_c$ emitted from the source. We consider this signal received to be $s_0(t)$
	\item The point scatterer will perturb the pressure field received by the receiver and we assume  this perturbed signal to be $s_p(t)$. 
	\item The point scatterer is assumed to be in the farfield and we model the scattering from it as a spherical source reflecting the energy to the receiver. 
\end{enumerate}
Under these assumptions,	
\begin{equation} \label{eqn34}
	s_m(t) = s_0(t) + s_p(t)
\end{equation}
Note that $s_0(t)$ can be computed with existing numerical simulation methods even for complex sources (as no motion is involved) and can be represented as $s_0(t) = s_0 e^{i\omega_c t}$. Our main focus will be on deriving an expression for $s_p(t)$  as it is the component of the signal relevant to ultrasound motion sensing. We utilize the analytical expressions derived earlier in the article.
We represent the scattering process in two parts:
\begin{enumerate}
	\item In the first step, the scatterer acts as a receiver and receives the incident field from the source. In this case, we use the expressions derived in section \ref{sec2}. 
	\item In the next step, the scatterer acts as a source and scatters a portion of the incident field back to the receiver with a reflection coefficient, $\alpha_p$. In this case, we use the expressions derived in section \ref{sec3}.
\end{enumerate}

Based on the above observations and from Eqns(\ref{eqn9} \& \ref{eqn22}),  $s_p(t)$ can be written as  
\begin{equation} \label{eqn35}
	s_p(t) =  p(\vec{r}_m,t) = \frac{\alpha_pp_0}{||\vec{r}_p(t)-\vec{r}_s||||\vec{r}_m - \vec{r}_p (t)||} \dfrac{1}{1-M_p \cos\left(\theta_p(t)\right)}  e^{i\omega_c \left( t - \dfrac{||\vec{r}_p(t) - \vec{r}_s||}{c}-\dfrac{1}{1-M_p \cos(\theta_p(t)) }\dfrac{||\vec{r}_m - \vec{r}_p (t)||}{c} \right)} 
\end{equation}

The analytic-signal representation of this yields

\begin{equation}\label{eqn36}
	A_m(t)=\frac{\alpha_pp_0}{||\vec{r}_p(t)-\vec{r}_s||||\vec{r}_m - \vec{r}_p (t)||} \dfrac{1}{1-M_p \cos\left(\theta_p(t)\right)} 
\end{equation}
\begin{equation}\label{eqn37}
	\phi_m(t)=\omega_c \left( t - \dfrac{||\vec{r}_p(t) - \vec{r}_s||}{c}-\dfrac{1}{1-M_p \cos(\theta_p(t)) }\dfrac{||\vec{r}_m - \vec{r}_p (t)||}{c} \right)
\end{equation}
\begin{equation} \label{eqn38}
	\omega(t)= \dfrac{d\phi(t)}{dt} = \omega_c - \vec{k}_{sp} .\vec{v}_p(t) + \dfrac{1}{1-M_p \cos(\theta_p(t)) } \vec{k}_{pm}. \vec{v}_{p}(t)
\end{equation}
where $\vec{k}_{sp} =\left(\dfrac{\omega_c}{c}\right) \dfrac{\vec{r}_p(t)-\vec{r}_s}{||\vec{r}_p(t)-\vec{r}_s||}$, $\vec{k}_{pm} =\left( \dfrac{\omega_c}{c}\right)\dfrac{\vec{r}_m-\vec{r}_p(t)}{||\vec{r}_m-\vec{r}_p(t)||}$,  $\vec{v}_p(t)=\frac{d\vec{r}_p(t)}{dt}$, $M_p = \dfrac{||\vec{v}_p ||}{c}$, and $\cos(\theta_p(t)) = \dfrac{(\vec{r}_m-\vec{r}_p)}{||\vec{r}_m-\vec{r}_p||}.\dfrac{\vec{v}_p} {||\vec{v}_p||}$. Here $\vec{k}_{sp}$ and $\vec{k}_{pm}$ are the instantaneous wave-vectors from the source to the target and target to the receiver respectively. 

\subsection{Some interesting observations}
\begin{enumerate}
	\vspace{12pt}
	\item \underline{Source and receiver are close and the target is in the far-field moving radially away from the source}:
	
	For this case, $\vec{k}_{sp}\approx-\vec{k}_{pm} = \vec{k}$ and $\theta_p(t)\approx180^{\circ}$, so 
	\begin{equation} \label{eqn38a}
		\omega(t)= \dfrac{d\phi(t)}{dt} = \omega_c -\vec{k}. \vec{v}_{p}(t) - \dfrac{1}{1+M_p}  \vec{k}. \vec{v}_{p}(t)  \approx  \omega_c -2\vec{k}. \vec{v}_{p}(t) \approx \omega_c\left(1-\dfrac{2||\vec{v}_p(t) ||}{c}\right)
	\end{equation}
	Note the factor of `2' multiplying $||\vec{v}_p(t) ||$ in the above expression. This is due to the point target first acting as a receiver and then as a source in the process of reflecting ultrasound to the receiver.
	
	In addition to the Doppler shift, since $||\vec{r}_p(t)-\vec{r}_s|| \approx ||\vec{r}_m - \vec{r}_p (t)||$, the amplitude of the signal for this case decays as $A_m(t) \approx (1/||\vec{r}_p(t)-\vec{r}_s||^2)$. In other words, the received signal strength drops by 6 dB for every 3 dB increase in distance to the source. This observation dictates the practical SNR limits for the maximum range of motion detection for ultrasound motion sensing applications.
	\vspace{12pt}
	\item \underline{Equation for the zero-Doppler shift }:
	From Eqn (\ref{eqn38}), we can get an equation for the zero Doppler-shift for arbitrary source and receiver locations by setting $\omega(t) - \omega_c=0$ i.e.,
	\begin{equation} \label{eqn38b}
- \vec{k}_{sp} .\vec{v}_p(t) + \dfrac{1}{1-M_p \cos(\theta_p(t)) } \vec{k}_{pm}. \vec{v}_{p}(t) = 0
	\end{equation}
	Noting that $\vec{k}_{sp} = \dfrac{\omega_c}{c} \vec{n}_{sp}$ and $\vec{k}_{pm}= \dfrac{\omega_c}{c} \vec{n}_{pm}$ where $\vec{n}_{sp} =\dfrac{\vec{r}_p(t)-\vec{r}_s}{||\vec{r}_p(t)-\vec{r}_s||}$, $\vec{n}_{pm} =\dfrac{\vec{r}_m(t)-\vec{r}_p}{||\vec{r}_p(t)-\vec{r}_s||}$, we can rewrite Eqn(\ref{eqn38b}) as
	\begin{equation} \label{eqn38c}
		- \vec{n}_{sp} .\vec{v}_p(t) + \dfrac{1}{1-\frac{1}{c}\vec{n}_{pm}.\vec{v}_p(t) } \vec{n}_{pm}. \vec{v}_{p}(t) = 0
	\end{equation}
	
	From Eqn(\ref{eqn38c}), we can see that the 
	\begin{itemize}
		\item Doppler-shift in frequency is not necessarily zero if the target is moving perpendicular to either source or receiver but is zero only when the target is moving at an angle given by the relation above. 
		
		\item Direction of motion for the zero doppler shift also depends on the location of the target relative to both source and receiver at that instant.
		 
	\end{itemize}
	
\end{enumerate}

Until now, we derived expressions for the signals in continuous-time without recourse to any temporal sampling. However, for numerical implementation of the simulation methodology, we need to utilize a discretized implementation as will be outlined later in section \ref{sec5.3}.  It is prefereable from a numerical computation standpoint that the signals $s_m(t)$ and $s_p(t)$ be represented in the baseband representation i.e., by removing the $e^{i\omega_ct}$ dependence in them. Note that while this is trivial from an analytical standpoint, numerical simulations benefit from this significantly as the target motion can now be sampled at much coarser time-scale without any temporal sampling constraints imposed by the ultrasonic wave frequency $\omega_c$. In fact, baseband signal processing is the key to several practical radar and ultrasound motion sensing applications. 

In the next section, we present the numerical method to compute the ultrasound motion sensing signal from a scattering target moving in a multi-reflection environment  with a prescribed motion trajectory.

\section{Point Scatterer moving in multi-reflection environment with a non-uniform source directivity: Integration with Image Source Method \label{sec5}}
In this section, we address four aspects of the numerical simulation that are key to extending the analytical methods presented above for numerical simulation of ultrasound motion sensing. These include
\begin{enumerate}
	\item Source Directivity
	\item Moving target motion representation 
	\item Temporal sampling and Baseband signal representation
	\item Image Source Method
\end{enumerate}

\subsection{Source directivity \label{sec5.1}}
The expressions derived up until now have focused on simple sources such as a planewave or point sources generating the pressure field. However, realistic sources typically have non-uniform directivity characteristics that need to be incorporated into the numerical simulation. Non-uniform directivity is the major factor that influences detection sensitivity and the Field-of-View of motion sensing technologies. In this section, we present details on how we repesent non-uniform source directivity used for the numerical simulations.

As mentioned earlier, we use geometrical acoustics for the numerical simulation. The sources used will be represented as a point source in space with an emission directivity function, $D(\theta, \phi)$, where $\theta$ is the azimuth angle and $\phi$ is the polar angle in the spherical polar coordinate system with center at the source location, $\vec{r}_s$. For such a source, the pressure field in an anechoic environment (i.e., free space and no reflections) is given by
\begin{equation} \label{eqn39}
	p(\vec{r}, t) = D(\theta, \phi) \dfrac{e^{-ik||\vec{r} - \vec{r}_s ||}} {|| \vec{r}  - \vec{r}_s||}
\end{equation}
Typically, these directivity functions are determined either by simulation or are measured for complex sources and will be used as the input for numerical simulation methodology presented here. For the representation in Eqn(\ref{eqn39}), $D(\theta, \phi)$ can be regarded as the pressure field measured at a unit distance (1 m) away from the source.
For a source with uniform directivity i.e., a spherically symmetric source,
\begin{equation} \label{eqn40}
D(\theta, \phi) = 1
\end{equation}
For a circular piston source, the directivity function is given by \cite{temkin1981elements}
\begin{equation} \label{eqn41}
	D(\theta, \phi) = \dfrac{2J_1\left (ka \sqrt{1-{(\hat{r}.\hat{n}_s)}^2} \right )} { ka\sqrt{1-{(\hat{r}.\hat{n}_s)}^2} }
\end{equation}
whre $\hat{r}$ is the unit-vector from the source location to the point of directivity measurement, $\hat{n}_s$ is the unit-normal to the piston surface, $a$ is the radius of the piston source,  $k=\frac{\omega_c}{c}$ is the wavenumber and $J_1$ is first-order Bessel function of first kind.

In this article, we use both cartesian and spherical-polar coordinates for numerical simulation and use the following mapping for unit vectors in cartesian to the spherical polar.
$\hat{x} : (\theta=0, \phi=90)$, $\hat{y}:  (\theta=90, \phi=90)$, and $\hat{z}: (\phi=0)$.

\begin{figure}[H]
	\centering
	\begin{subfigure}[t]{0.4\textwidth}
		\centering
		\includegraphics[ scale=0.15]{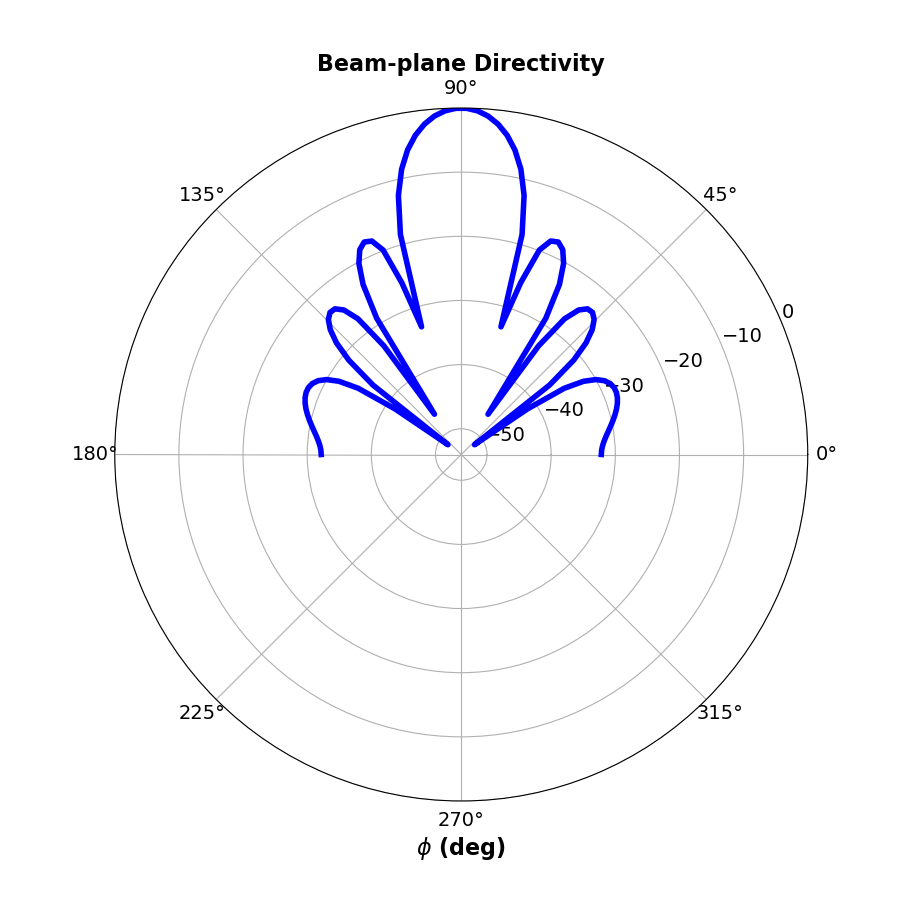}
		\caption{Beam plane directivity \label{fig1a}}
	\end{subfigure}
	\begin{subfigure}[t]{0.4\textwidth}
		\centering
		\includegraphics[scale=0.15]{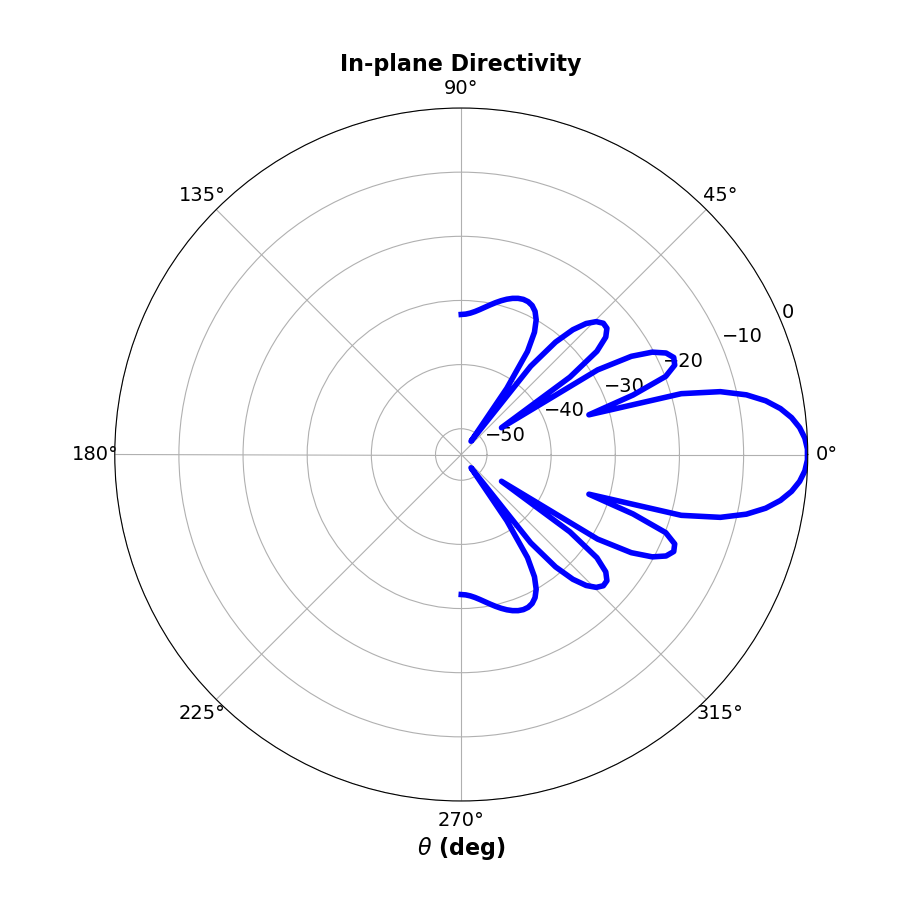}
		\caption{In-plane directivity \label{fig1b}}
	\end{subfigure}
	\caption{Piston source beam directivites for $n_s=(\theta=0^{\circ}, \phi=90^{\circ})$) \label{fig1}}
\end{figure}

\begin{figure}[H]
	\centering
	\begin{subfigure}[t]{0.4\textwidth}
		\centering
		\includegraphics[ scale=0.15]{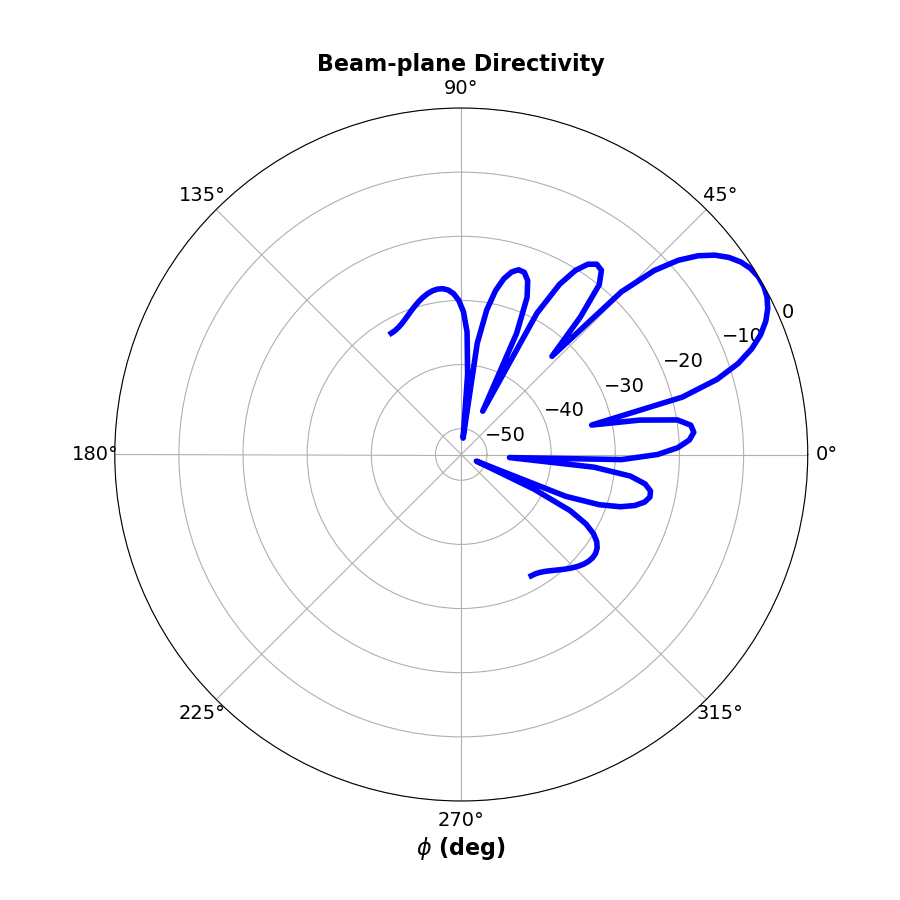}
		\caption{Beam plane directivity \label{fig2a}}
	\end{subfigure}
	\begin{subfigure}[t]{0.4\textwidth}
		\centering
		\includegraphics[scale=0.15]{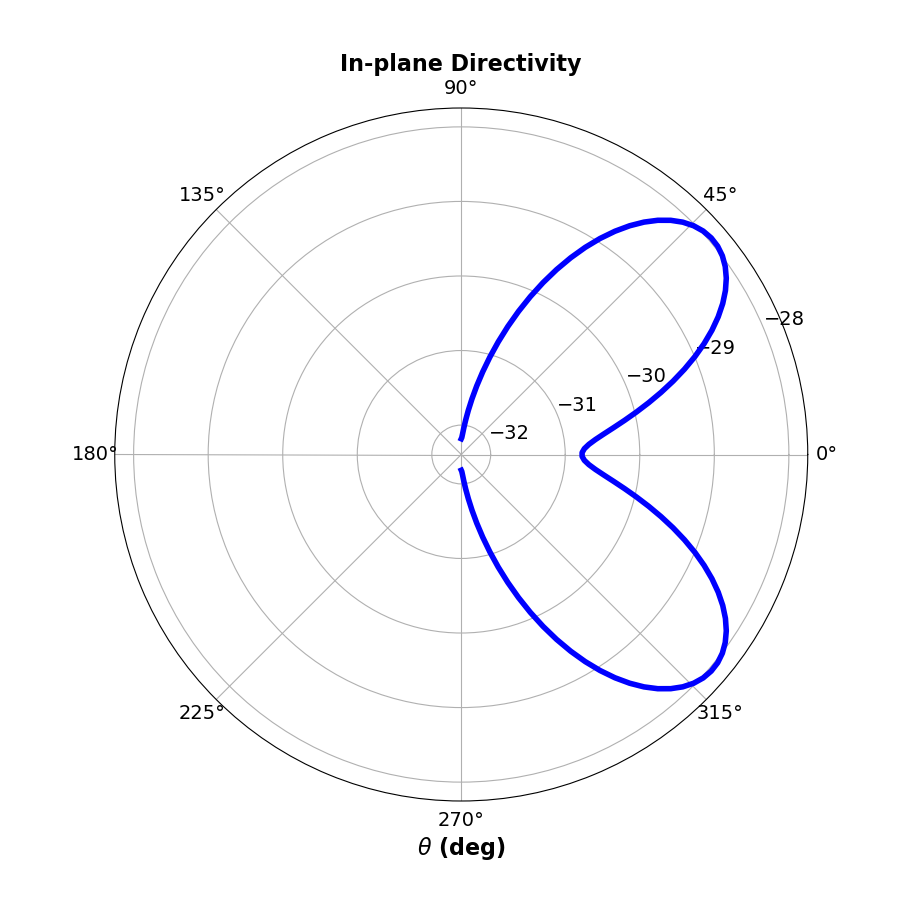}
		\caption{In-plane directivity \label{fig2b}}
	\end{subfigure}
	\caption{Piston source beam directivites for $n_s=(\theta=0^{\circ}, \phi=30^{\circ})$ \label{fig2}}
\end{figure}

Figures (\ref{fig1} \& \ref{fig2}) show the beam directivities plotted for a circular piston source ($a = 20$ cm) directed in x-direction and for the case it is directed at $\phi=30^{\circ}$ from the z-axis. For each case, we show two directivity plots in dB-scale:
\begin{enumerate}
	\item  Beam-plane directivity: It is the directivity in the plane of the beam's main lobe.
	\item In-plane directivity: It is the directivity in the x-y plane through the source. 
\end{enumerate}

Observe that in Figure \ref{fig2b}, the in-plane directivity shows two lobes which actually are the side lobes from the source intersecting the x-y plane. Hence, the field strength shown in the plot is lower than that in the beam plane in Figure \ref{fig2a}.

\subsection{Moving target motion representation \label{sec5.2}}
For the simulations presented in this article, we consider planar targets represented by point-clouds. Let $\mathcal{P}_0$ denote the set of points in the point cloud at $t=0$. We consider only the rigid body translational motions in this article. For such case, motion of the target is represented by translation of one of the points in $\mathcal{P}_0$. Let $\mathcal{P}_t$ denote the collection of points at the time-instant `$t$'.
If $\vec{v}(t)$ denotes the velocity profile for the translation motion, then for each $\vec{r}_{p}$ in $\mathcal{P}_0$, we have 
\begin{equation} \label{eqn42}
	\vec{r}_p(t) = \vec{r}_{p}(t=0) + \int_{0}^{t}\vec{v}(t') dt'
\end{equation}
Equation (\ref{eqn42}) will be used for determining the location of points in  $\mathcal{P}_t$. Note that rigid body rotations can also be easily incorporated in this framework by including a rotation-term in Eqn(\ref{eqn42}). 
For our simulations, we consider a stepping speed profile which is defined by 
\begin{equation}\label{eqn42a}
	||\vec{v}(t)|| = v(t)= \begin{cases}
		v_{min}+(v_{max}-v_{min} )\left(\dfrac{2t}{T_{step}}\right)^2 & 0 \leq t \leq \frac{T_{step}}{2}\\
		v_{max}-(v_{max}-v_{min} )\left(\dfrac{2\left(t-\frac{T_{step}}{2}\right)}{T_{step}}\right)^2 & \frac{T_{step}}{2} \leq t \leq T_{step}\\
		v(t-T_{step}) &  t >T_{step}
	\end{cases}  
\end{equation}
The average speed for such a speed profile is $v_{avg}=\frac{2v_{min}+v_{max}}{3}$. For a constant speed motion, we set $v_{min} = v_{max} = v_{avg}$. Also, we set $T_{step} = 1 \hspace{2 pt} sec$ for all the simulations presented in the article.

With the above target representation,  the scattered field from the target recevied at receiver will be the sum of radiated field from point sources located in the point cloud  and given by
\begin{equation} \label{eqn43}
	s_m(t) = \int_{ \vec{r}_p \in \mathcal{P}_t} p_m(t; \vec{r}_p) dA_p
\end{equation}
where $p_m(t; \vec{r}_p)$ is the pressure field at the receiver due to the scattered emission from the point $\vec{r}_p$ computed by the methods described in section \ref{sec4}. Note that the integration is performed over all the points in the target and $dA_p$ is the target elemental area surrounding the point.

\subsection{Temporal sampling and Baseband signal representation \label{sec5.3}}
In subsection above, we described the velocity profile we use for motion trajectories. The expressions are presented in continuous time i.e., there is no discretization specified for the time. However, to implement the above methodology numerically, we need to sample the motion at discretized time-steps with a step-size, $\Delta t_{motion}$. Analogously, the received signal needs to be sampled at discrete time steps keeping in mind that the expressions for the pressure field  $p_m(t; \vec{r}_p)$ in Eqn (\ref{eqn43}) involves sinusoidal component,  $e^{i\omega_ct}$ with carrier frequency $\omega_c$. For example,  if $\omega_c=33$ kHz, to adhere to the Nyquist limit, we need to sample the sinusoidal signal at atleast $\Delta t_{wave} =0.015 \hspace {2 pt} ms$.   If we adhere to this limit, 1 sec of motion simulation would require $\approx 66000$ time steps which is practically infeasible for simulating longer motion durations. However, the baseband implementation overcomes this by eliminating the need for sampling the ultrasonic wave and thereby significantly reducing the computational burden for running these simulations. We discuss the rationale for this by considering the expression from Eqn(\ref{eqn9}) and is equally valid for all other analytical expressions in this article. We rewrite Eqn(\ref{eqn9}) by eliminating the term,  $e^{i\omega_ct}$ that requires higher temporal sampling as 
\begin{equation} \label{eqn43a}
	p_{baseband}(\vec{r},t) = \frac{p_0}{||\vec{r}-\vec{r}_s||}e^{-i\omega_c \frac{||\vec{r}-\vec{r}_s||}{c}}
\end{equation}
 Note that the above expression eliminates any explicit time dependence and thereby the need for high temporal sampling while still retaining all the motion related components (positions and velocities) required for the numerical simulation. So, in effect, the numerical simulation can be implemented by using baseband versions of all the expressions derived earlier with no loss of motion signatures in the resultant outputs. We need only one temporal discretization time-step, $\Delta t_{motion}$ and  use the same for discretizing/sampling the baseband signal. We use $\Delta t_{motion} = 0.5$ ms for all the results presented in this article.  
 
With this observation, we rewrite expressions for the discretized version of received signal in Eqn(\ref{eqn43}) at $n^{th}$ time-step as 
\begin{equation} \label{eqn43b}
 	s_m(t_{n}) = \int_{ \vec{r}_p \in \mathcal{P}_{t_{n}}} p_{m}(t_n; \vec{r}_p(t_{n})) dA_p
\end{equation}
where $t_{n} = n \Delta t_{motion}$ and $p_{m}(t_n; \vec{r}_p(t_{n}))$ is the baseband version of the corresponding quantity. The discrete version of the baseband signal at the receiver is given by $s_m(t_{n})$ and the Short Time Fourier Transform (STFT) of this basbeband signal will show all the motion related signatures associated with the target in motion as will be demonstrated later through several examples.  

\subsection{Image Source Method \label{sec5.4}}
Image Source Method (ISM) \cite{allen1979image,vorlander1989simulation}  is a widely used method in room-acoustics for computing the Room Impulse Responses. The implementation involves treating the reflections in the environment as though they were from an array of virtual sources that are images of the real source reflected about the room walls. Each image source has a reduced strength corresponding to the order of reflections it went through and the corresponding reflection coefficients of the walls. ISM is a very accurate method particularly for the early reflections in the multi-reflection environment and has been widely used in geometrical acoustics \cite{savioja1999modeling,koutsouris2013combination}. Since ultrasound decays much faster than the audible sound due to the attenuation ($\approx$ 0.7-1dB/m at 30-40 kHz), accurate modeling of early reflection suffices for airborne ultrasound based motion sensing. 

In this article, we use the ISM method for computing the incident field on the target by computing the pressure field at each of the points in $\mathcal{P}_t$ as the target is moving in the environment by summing up the field contributions from each of the  image-sources. Each of the image source is a point source with the directivity function and so the expressions dervied in section \ref{sec4} can be applied to each of the image sources. If $\mathcal{S}$ denotes the set of all real and image sources, then the received signal at the receiver is given by 
\begin{equation} \label{eqn44}
	s_m(t) = \sum_{\mathcal{S}_i \in \mathcal{S}}  \left(\int_{ \vec{r}_p \in \mathcal{P}_t} p_m(t; \vec{r}_p, \mathcal{S}_i) dA_p \right)
\end{equation} 
where $p_m(t; \vec{r}_p, \mathcal{S}_i)$ the received signal due to emission from the image-source $\mathcal{S}_i$ scattered from the point, $\vec{r}_p$, to the receiver. Note that the strength of the image sources depends on the order of the reflection being considered and it is incorporated when computing  $p_m(t; \vec{r}_p, \mathcal{S}_i)$ for each $\mathcal{S}_i$.

\section{Results \label{sec6}}
In this section, we present results from the numerical simulation methodology presented above. We consider the room with dimensions $6 m  \times 5 m  \times 3 m$ for all the simulations presented in this section. The parameters used for the simulation setup are shown in the table below:

\begin{table}[H]	
	\centering
	\caption{Table showing the parameters used for the simulation. \label{tab1}}
\begin{tabular}{c c}
	\hline\hline
	Source Location & (0, 0, 0)  \\
	Source radius (m) & 0.02 \\
	Receiver location & (0, 0, 0.05 m) \\
	Carrier frequency & 33000 Hz \\
	Ultrasonic attenuation in air & 1 dB/m \\
	Target Reflection coefficient & $\alpha_p$ = 1 \\
	Wall 1 & x=-1 m \\
	Wall 2 & x=5 m \\
	Wall 3 & y=-2.5 m\\
	Wall 4 & y = 2.5 m\\
	Wall 5  & z =1.8 m \\
	Wall 6 & z=-1.2 m \\
\end{tabular}

\end{table}

For all the results presented in the article, we discretize the target motion velocity profiles with a sampling frequency of 2 kHz i.e., with a temporal sampling of $\Delta t _{motion}= 0.5 \hspace{2pt} ms$.

\subsection{Point target in motion}
In this section, we consider a point target in motion i.e., a point cloud with just one point. These examples are chosen to illustrate how the simulation methodology presented above captures the key features of the signals used for Doppler based motion sensing. In this regard, we present illustrative examples that depict the influence of source directivity, motion trajectory, velocity profiles, and the influence of wall reflections on the Doppler signal characteristics at the receiver. We depict the effect of each separately to better illustrate the findings. 

\subsubsection{Influence of velocity profile}
\begin{figure}[H]
	\centering
		\begin{subfigure}[t]{0.4\textwidth}
		\centering
		\includegraphics[scale=0.15]{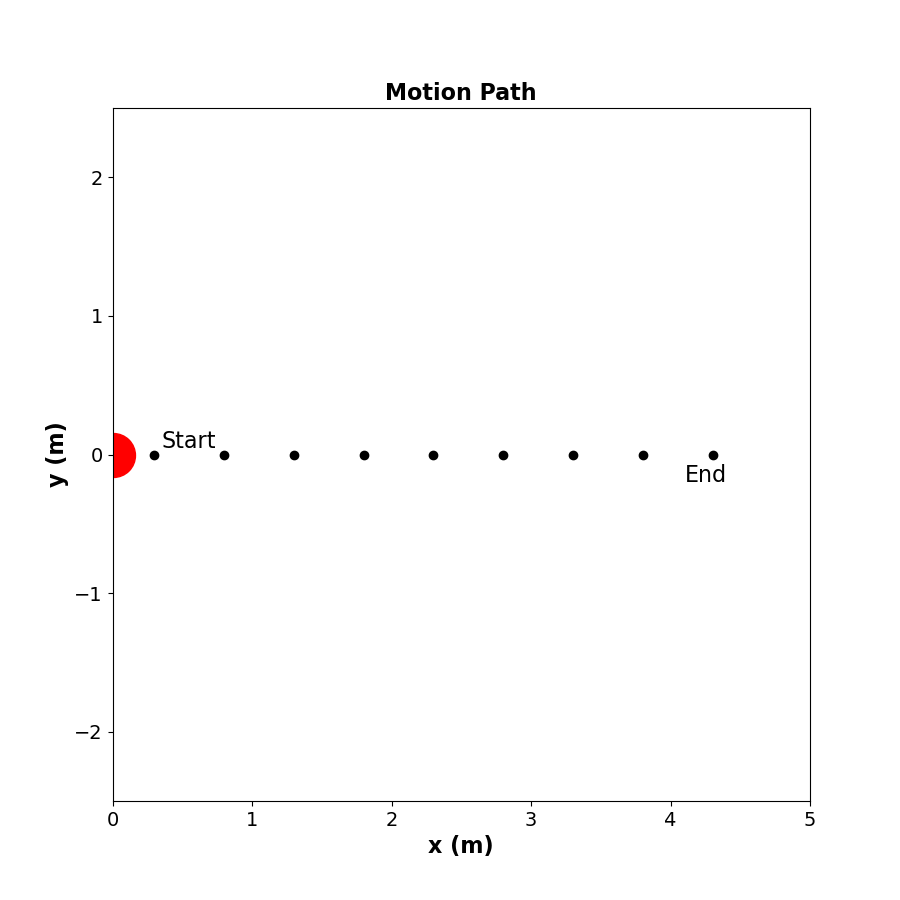}
		\caption{\label{fig3a}}
	\end{subfigure}
	\begin{subfigure}[t]{0.4\textwidth}
		\centering
		\includegraphics[scale=0.15]{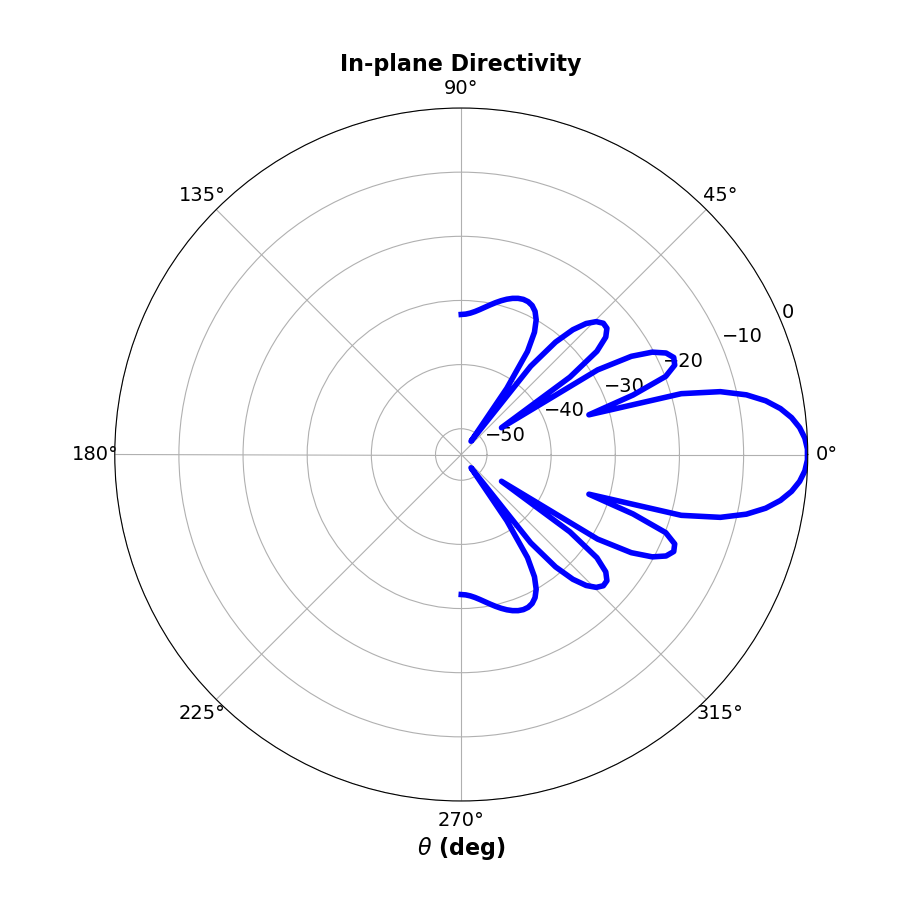}
		\caption{\label{fig3b}}
	\end{subfigure}
	\begin{subfigure}[t]{0.4\textwidth}
		\centering
		\includegraphics[scale=0.15]{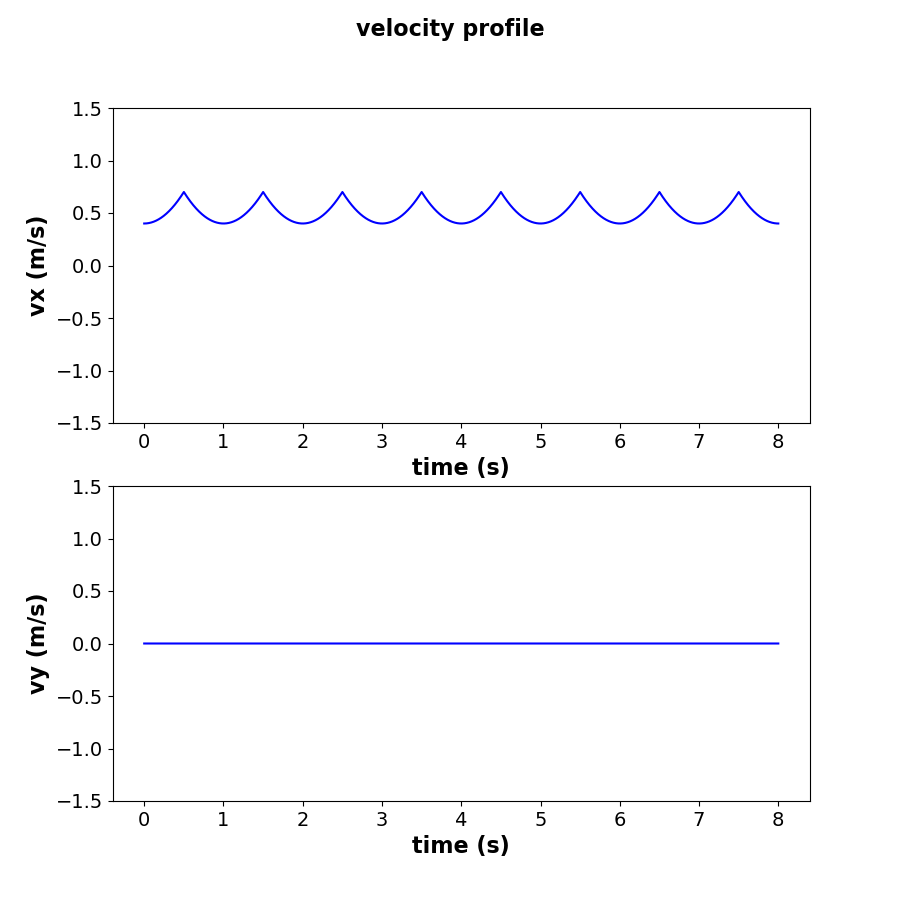}
		\caption{\label{fig3c}}
	\end{subfigure}
	\begin{subfigure}[t]{0.4\textwidth}
		\centering
		\includegraphics[scale=0.14]{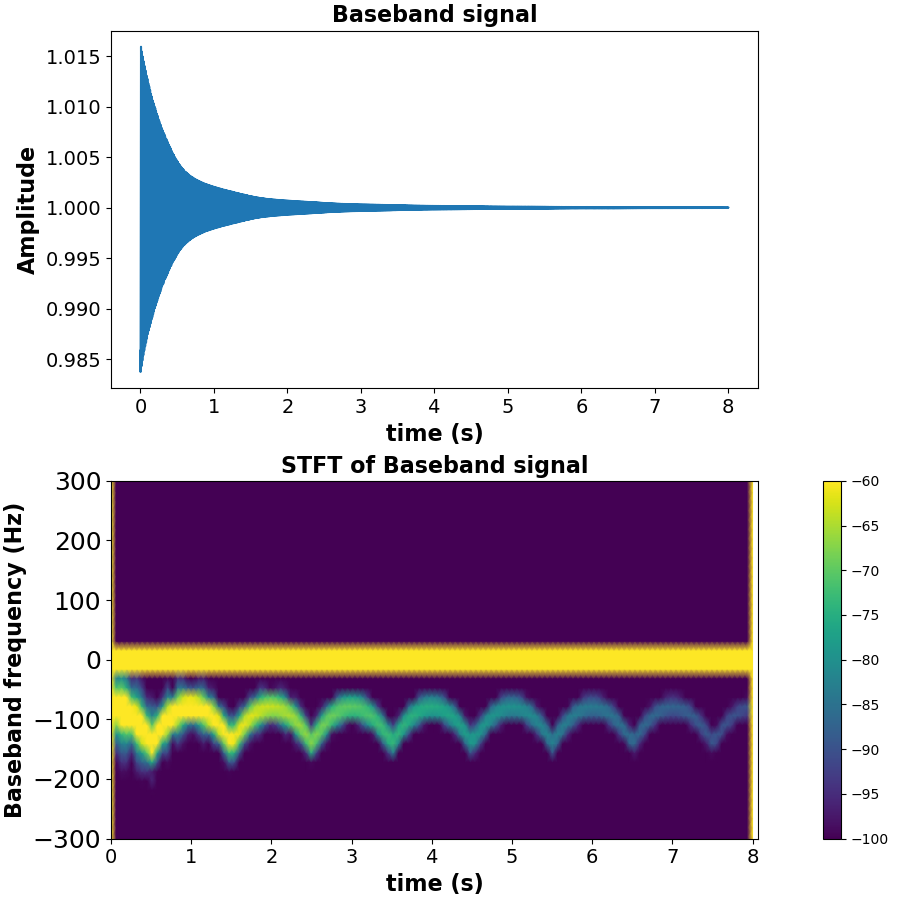}
		\caption{\label{fig3d}}
	\end{subfigure}
	\begin{subfigure}[t]{0.4\textwidth}
		\centering
		\includegraphics[scale=0.15]{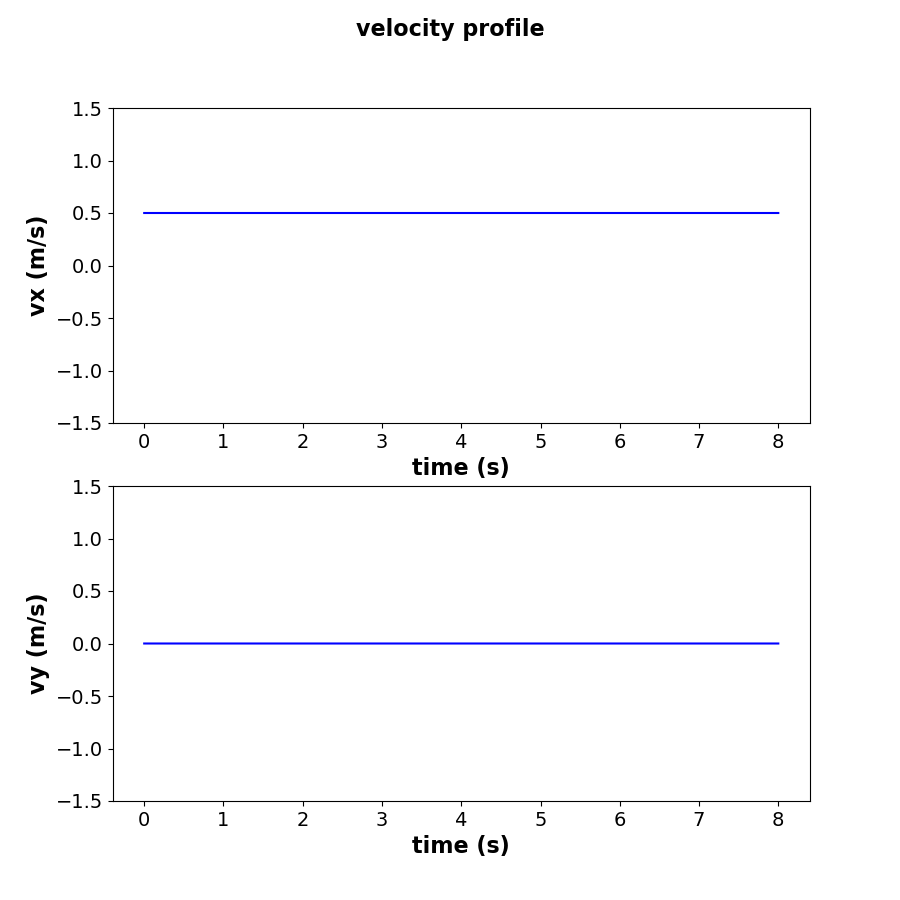}
		\caption{\label{fig3e}}
	\end{subfigure}
	\begin{subfigure}[t]{0.4\textwidth}
		\centering
		\includegraphics[scale=0.14]{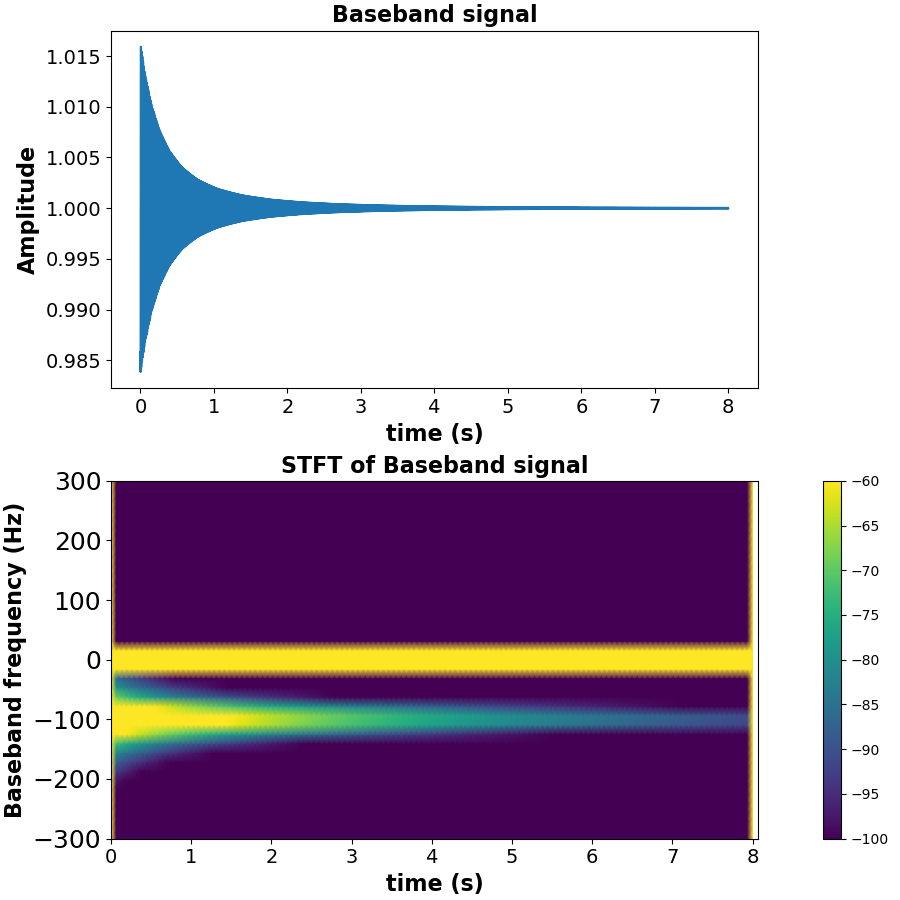}
		\caption{ \label{fig3f}}
	\end{subfigure}
	\caption{(a) On-axis Motion Trajectory (b) Source in-plane directivity (c) Stepping velocity Profile (d) Baseband signal and Doppler characteristics for stepping velocity profile (e) Constant velocity profile (f) Baseband signal and Doppler characteristics for a constant velocity profile}
\end{figure}

Here, we present and compare the results for two different velocity profiles---stepping profile with $v_{min} = 0.4 \hspace{2pt} m/s$, $v_{max} = 0.7 \hspace{2pt} m/s$, and motion at a constant speed, $v_{min} = v_{max} = 0.5 \hspace{2pt} m/s$ in Eqn(\ref{eqn42a}). The velocities of both profiles are set so that the average speed is the same for both the velocity profiles i.e., $v_{avg}=0.5$. In other words, the trajectory and the distance covered by the target is the same during the motion for both the cases. We consider the target to be moving away from the source along the x-axis as depicted in the motion trajectory  plot in Figure \ref{fig3a}.  The target starts at 0.3 m from the source and moves away from the source for 8 sec. The ``Start" and ``End" points of the motion trajectories are labeled in the figure.  The source used for this simulation is a circular piston source emitting along the x-axis as shown in the Figure \ref{fig3b}. The velocity profiles we use for comparing the Doppler characteristics are shown in Figures \ref{fig3c} and \ref{fig3e} for the stepping and constant speed motions respectively. The baseband signals and the corresponding Short-Time-Fourier-Transforms (STFT) are shown in Figures \ref{fig3d} and \ref{fig3f} respectively. Clearly, the STFT shows different doppler characteristics even though the average speed for both the motion profiles are the same. For the case of constant velocity profile, the Doppler frequency shift  observed is 100 Hz that is in agreement with the expression derived in Eqn(\ref{eqn38a}) i.e., $\left( -\frac{2*|| \vec{v}_p||}{c} f_c \right) =  \left(-\frac{2*0.5}{330} * 33000\right) = -100 \textrm{ Hz}$. For the case of stepping speed profile, we see the Doppler frequency oscillates between a minimum and maximum corresponfing to the speeds $v_{min}$ and $v_{max}$.  In addition, one can clearly see the decay of the baseband signal strength as the target moves away from the source in both the cases. This example illustrates that the simulation methodology captures both Doppler and the signal strength characteristics that are key to ultrasound based motion sensing methodologies. 
 
\subsubsection{Influence of motion trajectory}
\begin{figure}[H]
	\centering
	\begin{subfigure}[t]{0.4\textwidth}
		\centering
		\includegraphics[scale=0.15]{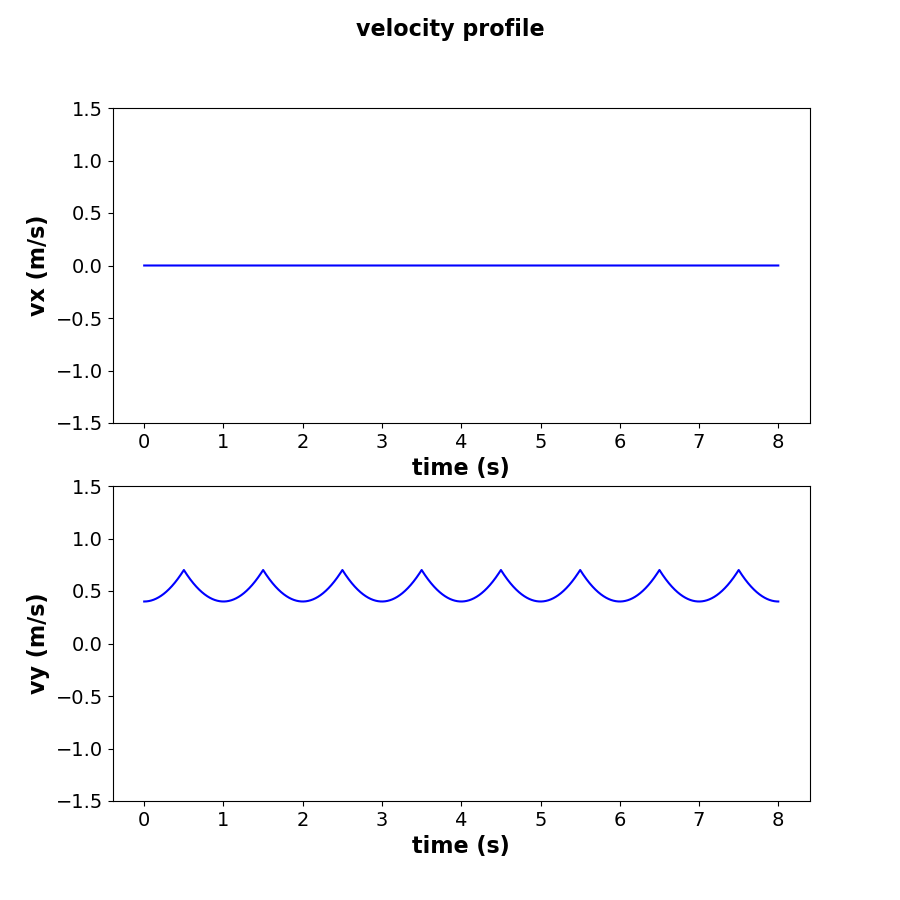}
		\caption{\label{fig4a}}
	\end{subfigure}
	\begin{subfigure}[t]{0.4\textwidth}
	\centering
	\includegraphics[scale=0.15]{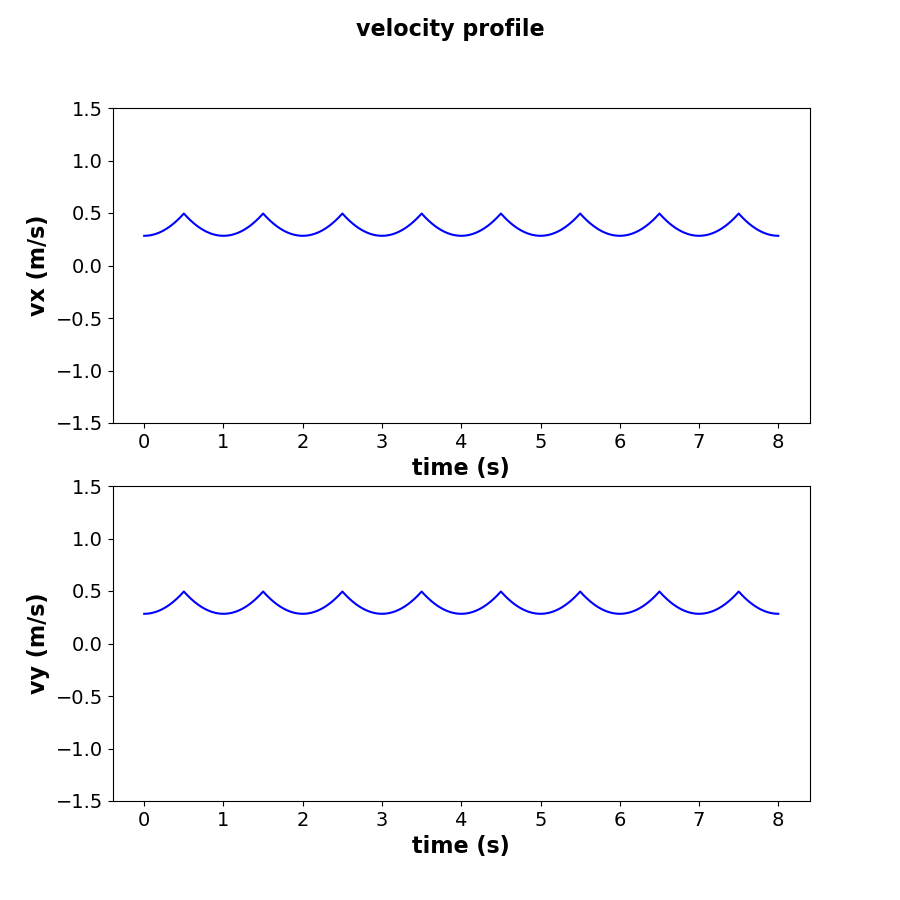}
	\caption{\label{fig4b}}
    \end{subfigure}
	\begin{subfigure}[t]{0.4\textwidth}
		\centering
		\includegraphics[scale=0.15]{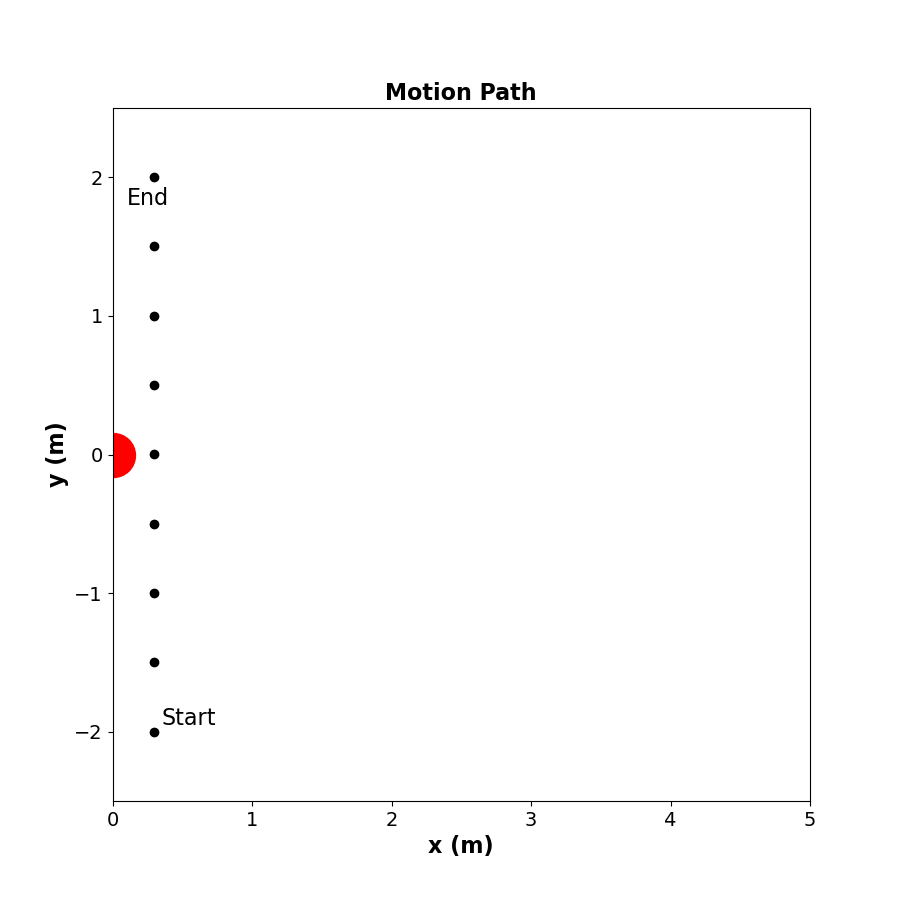}
		\caption{\label{fig4c}}
	\end{subfigure}
	\begin{subfigure}[t]{0.4\textwidth}
		\centering
		\includegraphics[scale=0.15]{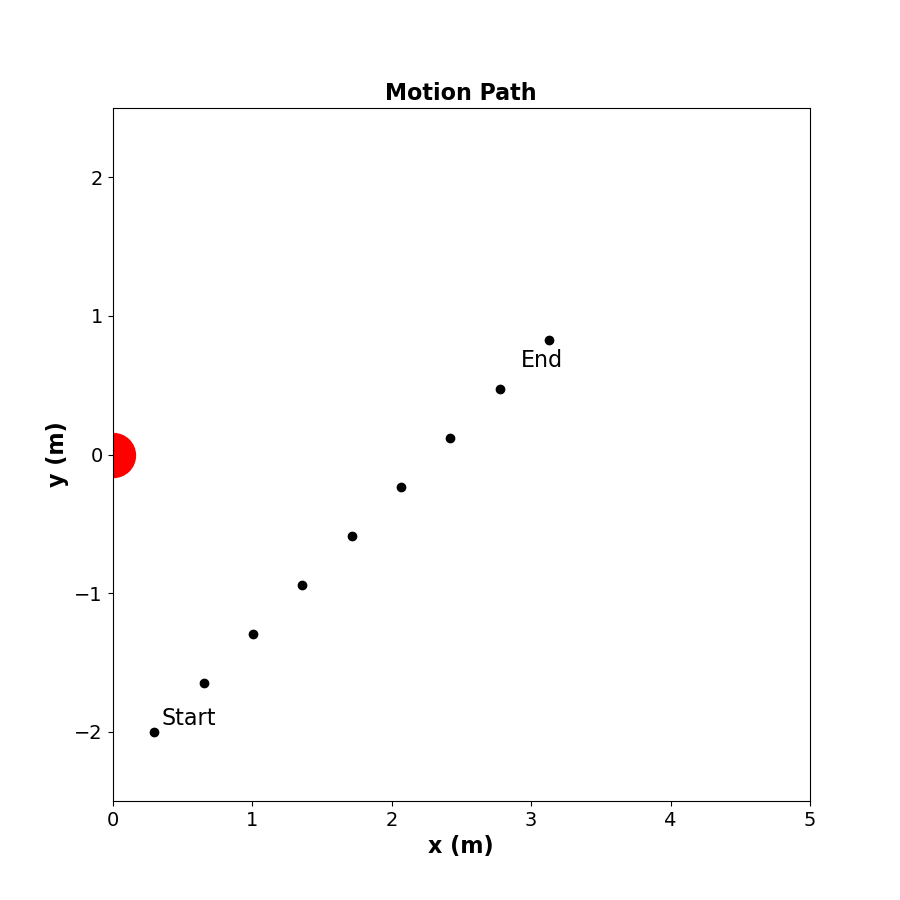}
		\caption{\label{fig4d}}
	\end{subfigure}
	\begin{subfigure}[t]{0.4\textwidth}
		\centering
		\includegraphics[scale=0.15]{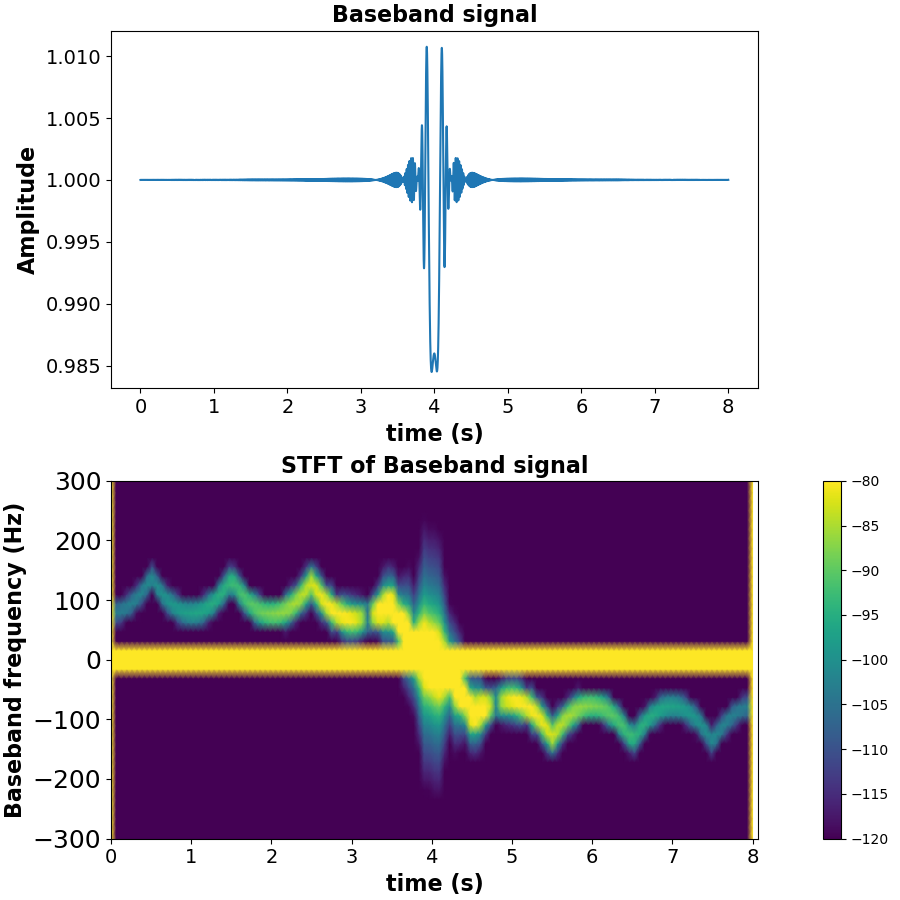}
		\caption{\label{fig4e}}
	\end{subfigure}
	\begin{subfigure}[t]{0.4\textwidth}
		\centering
		\includegraphics[scale=0.15]{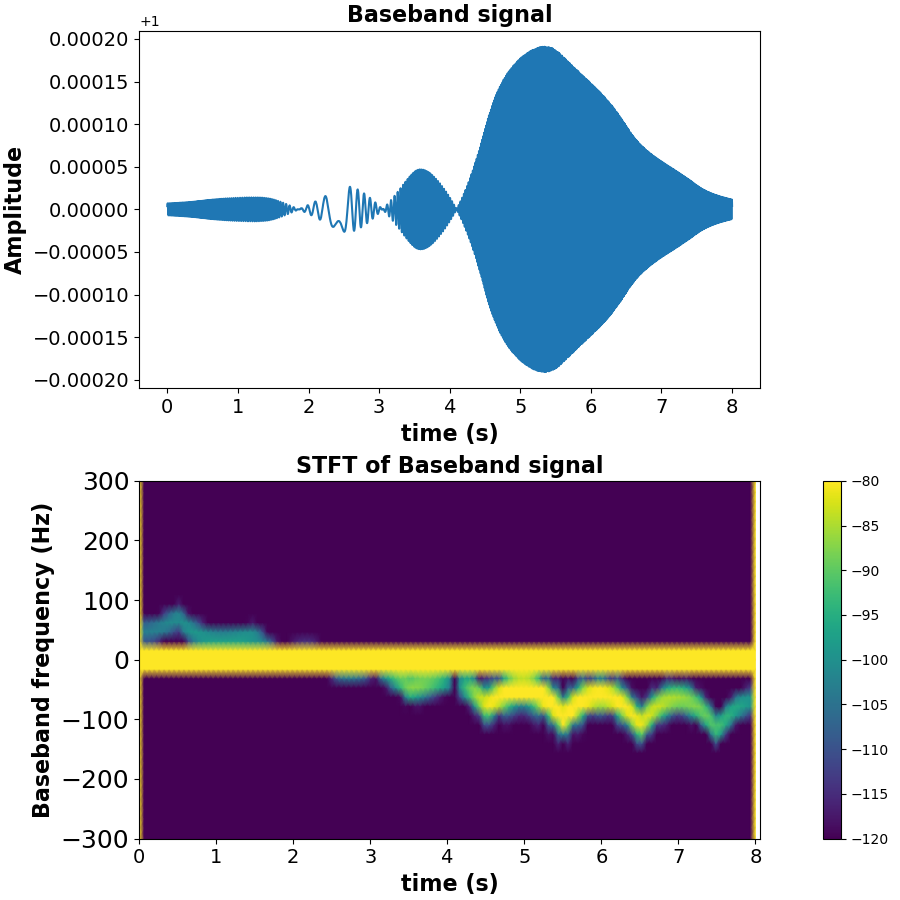}
		\caption{\label{fig4f}}
	\end{subfigure}
	
	\caption{(a) Transverse motion velocity profile(b) Diagonal motion velocity profile (c) Transverse motion trajectory (d) Diagonal motion trajectory (e) Baseband signal and Doppler characteristics for transverse motion trajectory  (f) Baseband signal and Doppler characteristics for diagonal motion trajectory}
\end{figure}

Here, we illustrate the influence of the motion trajectory on the signal characteristics observed in the simulation. The source is still assumed to be a circular piston source pointing in the x-direction. We choose two different  trajectories with the same motion speed profiles. Figures \ref{fig4a} and \ref{fig4b} show the velocity profiles used and Figures \ref{fig4c} and \ref{fig4d} show the corresponding motion trajectory. The target starts from the same location (x=0.3 m, y=-2 m, z=0) and moves along different paths with the same speed---one path where the target is moving transverse to the source and the other path where the target is moving at an angle of 45 degrees to the y-axis. Figures \ref{fig4e} and \ref{fig4f} depict the baseband signals and their respective STFT's for each of the cases. For the case of motion transverse to the source, one can clearly see the baseband signal is much lower in strength to begin with and as the target approaches the main-lobe of the source directivity which is along the x-axis (see Figure \ref{fig3b}), we get a stronger signal. At the same time, since the target is moving transverse to the source, the Doppler frequency transitions from a positive value to zero and then to a negative value. Note that this transition is symmetric as the motion and source directivity are symmetric for this motion trajectory. On the other hand, Figure \ref{fig4f} shows the result for the case where the target is moving at 45 degrees where the motion trajectory is asymmetric with respect to the main-lobe of the source directivity. Clearly, the signal characteristics of the baseband signal (Top panel of Figure \ref{fig4f}) shows asymmetric response along the motion trajectory. In fact, it is not monotonic because the user motion trajectory sequentially goes through the side-lobes, nulls, and the main-lobe where the signal gets the strongest. In this case, since the target is moving at a speed of 0.5 m/sec, the  motion trajectory intersects the main-lobe (i.e., x-axis) at $(2.3 , 0 , 0 )\hspace{2pt} m$ which the target reaches at $\frac{\sqrt{(2.3-0.3)^2 + (0- (-2))^2)}}{0.5} = 5.6 \hspace{2pt} sec$ where the signal strength is the strongest validating the simulation methodology. Also, note that the Doppler shift in \ref{fig4f} (Bottom panel) transitions from positive to zero to negative as the target is first approaching and then receding from the source. The point of closest approach could be calculated for this motion trajectory to be $(1.15, -1.15, 0) \hspace{2pt} m$ which the target reaches at around $\frac{\sqrt{(1.15-0.3)^2 + (-1.15- (-2))^2)}}{0.5} = 2.4 \hspace{2pt} sec$ which is where the Doppler frequency transitions through zero in Figure \ref{fig4f}. This example depicts the capability of the proposed methodology to capture the signal characteristics for different motion trajectories.

\subsubsection{Influence of reflections}
\begin{figure}[H]
	\centering
	\begin{subfigure}[t]{0.4\textwidth}
		\centering
		\includegraphics[scale=0.15]{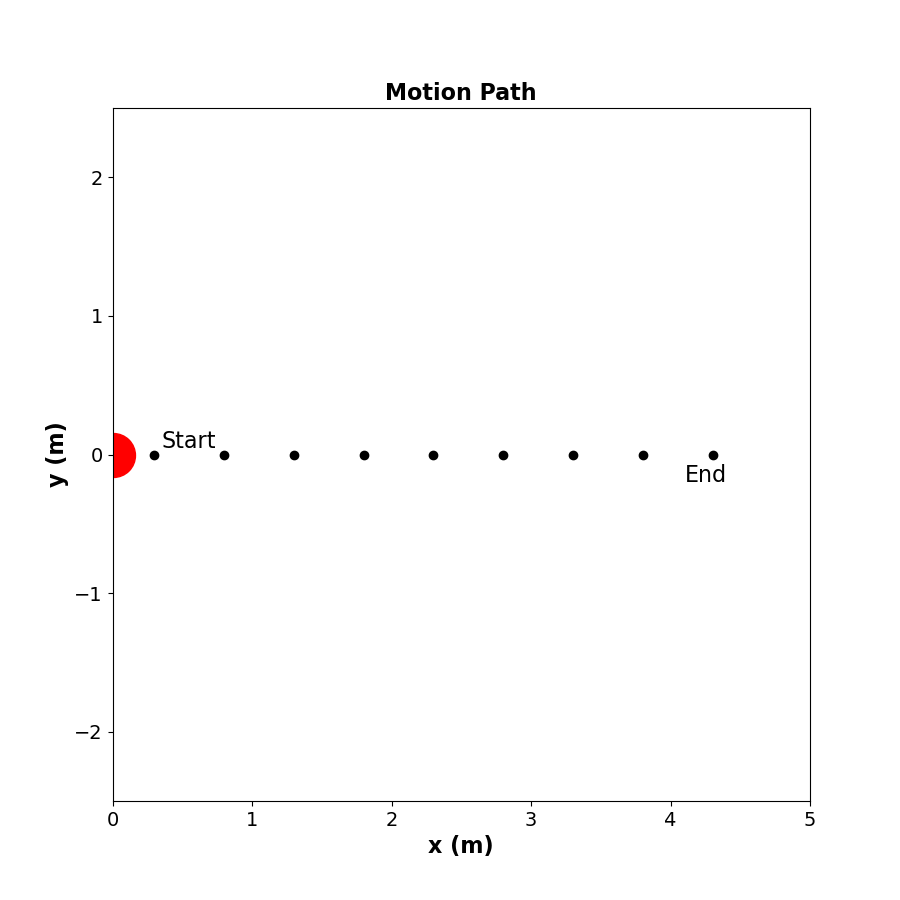}
		\caption{\label{fig5a}}
	\end{subfigure}
	
	\begin{subfigure}[t]{0.4\textwidth}
		\centering
		\includegraphics[scale=0.15]{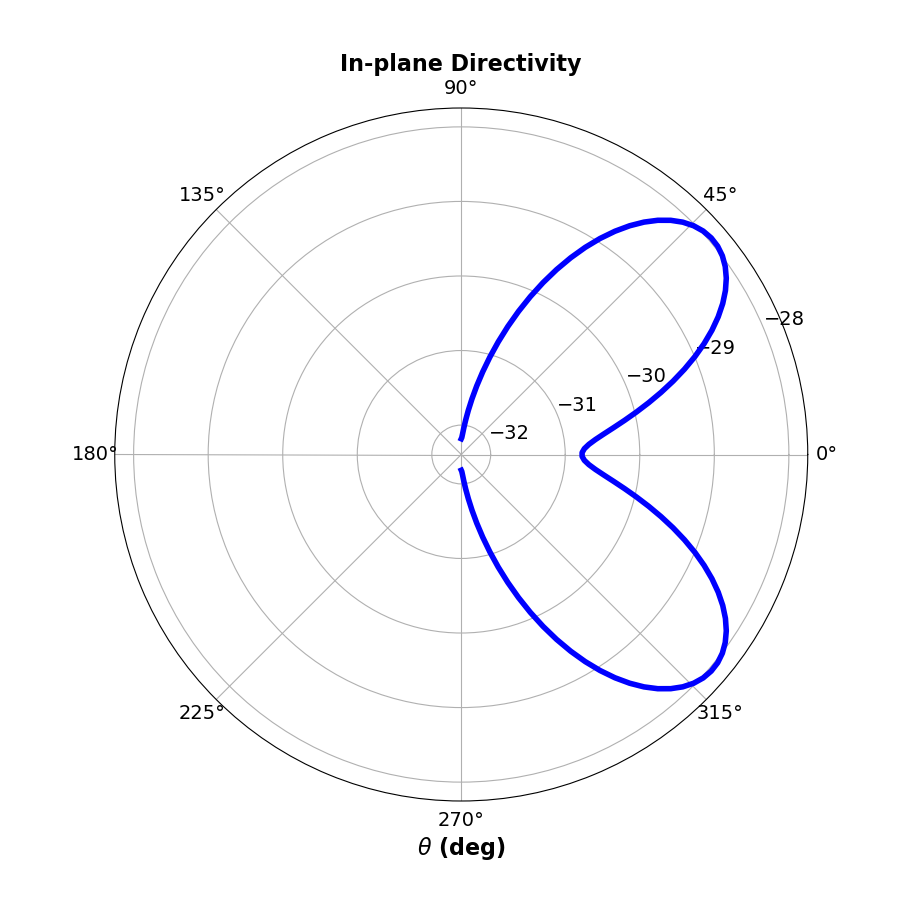}
		\caption{\label{fig5b}}
	\end{subfigure}
	\begin{subfigure}[t]{0.4\textwidth}
		\centering
		\includegraphics[scale=0.15]{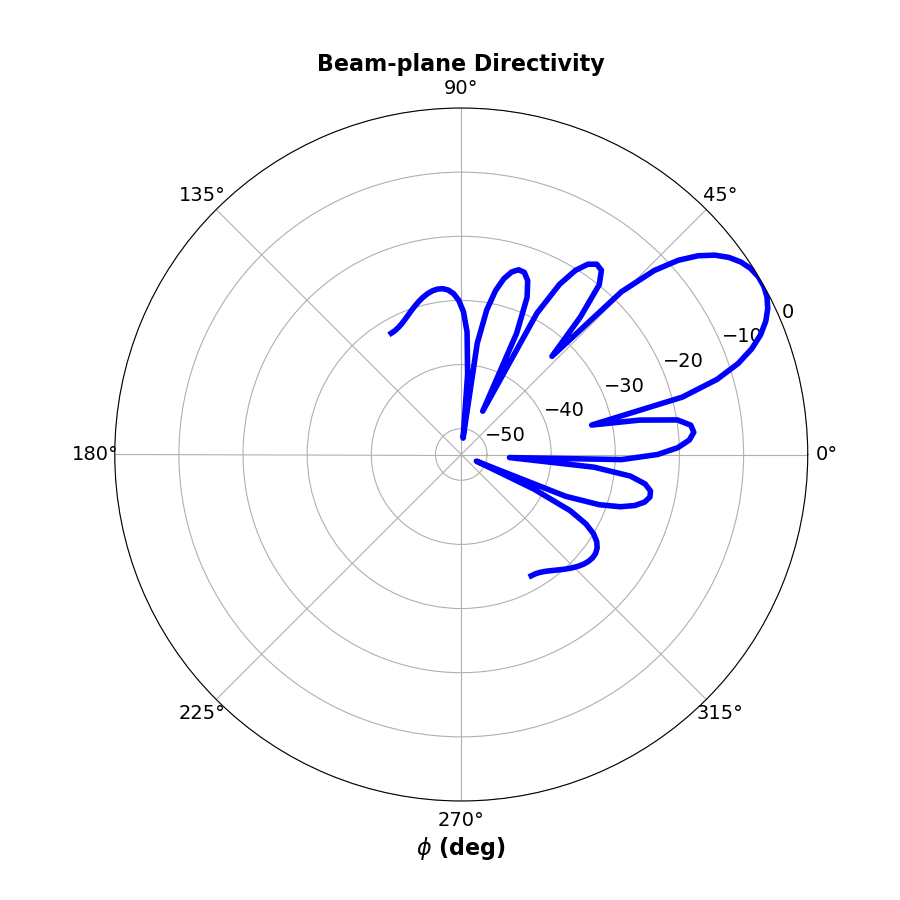}
		\caption{\label{fig5c}}
	\end{subfigure}
		\begin{subfigure}[t]{0.4\textwidth}
		\centering
		\includegraphics[scale=0.15]{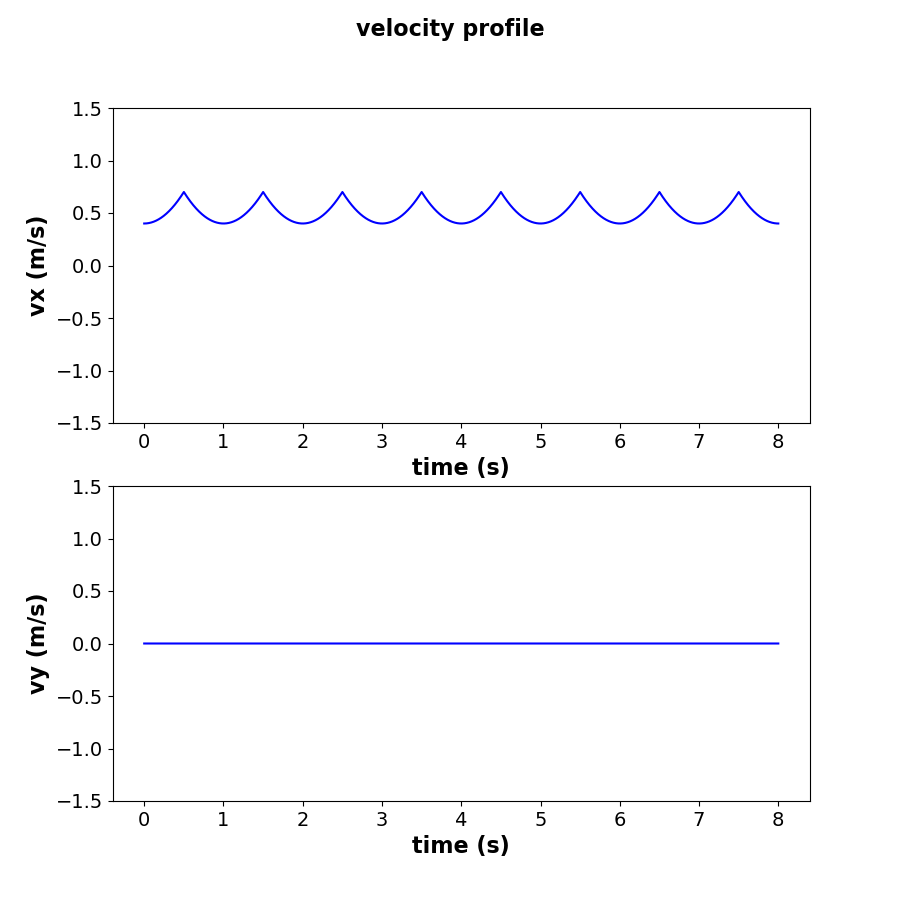}
		\caption{\label{fig5d}}
	\end{subfigure}
	\begin{subfigure}[t]{0.4\textwidth}
		\centering
		\includegraphics[scale=0.15]{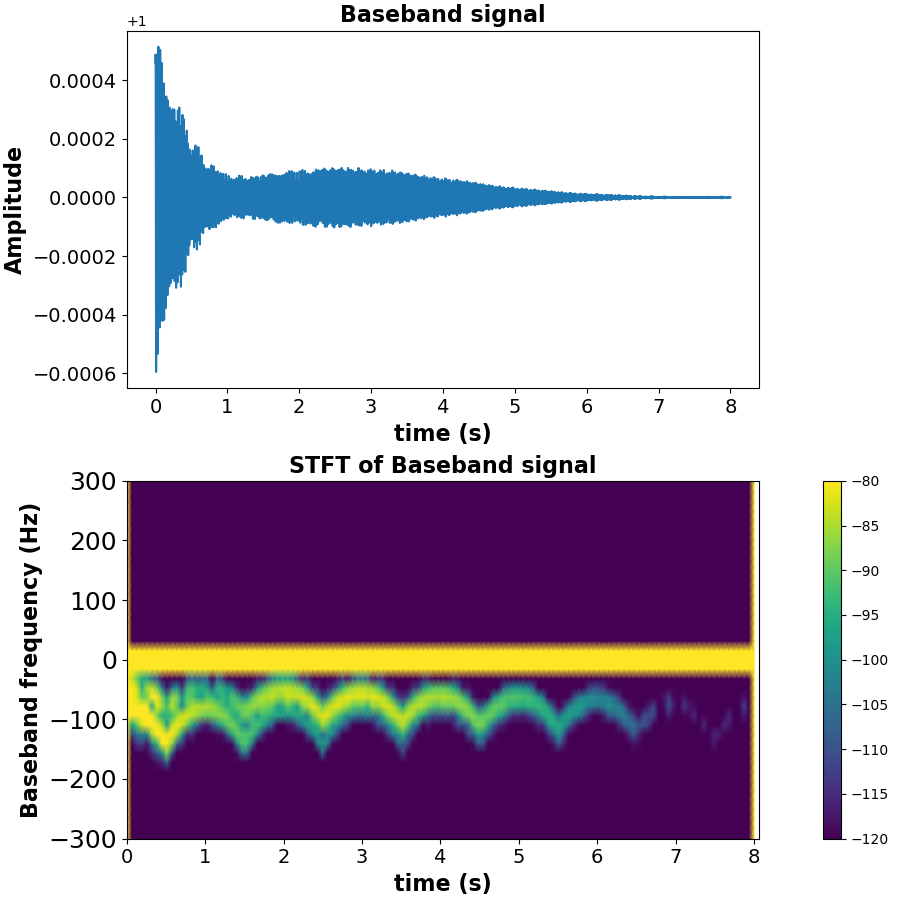} 
		\caption{\label{fig5e}}
	\end{subfigure}
	\begin{subfigure}[t]{0.4\textwidth}
		\centering
		\includegraphics[scale=0.15]{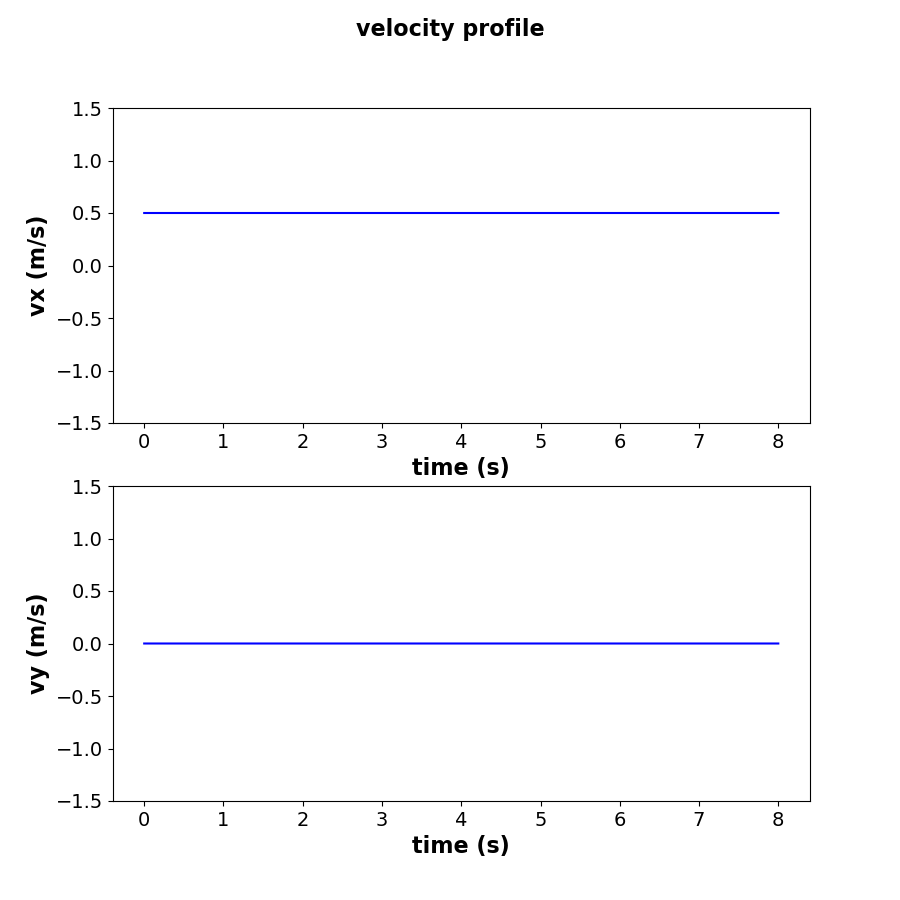}
		\caption{\label{fig5f}}
	\end{subfigure}
	\begin{subfigure}[t]{0.4\textwidth}
		\centering
		\includegraphics[scale=0.15]{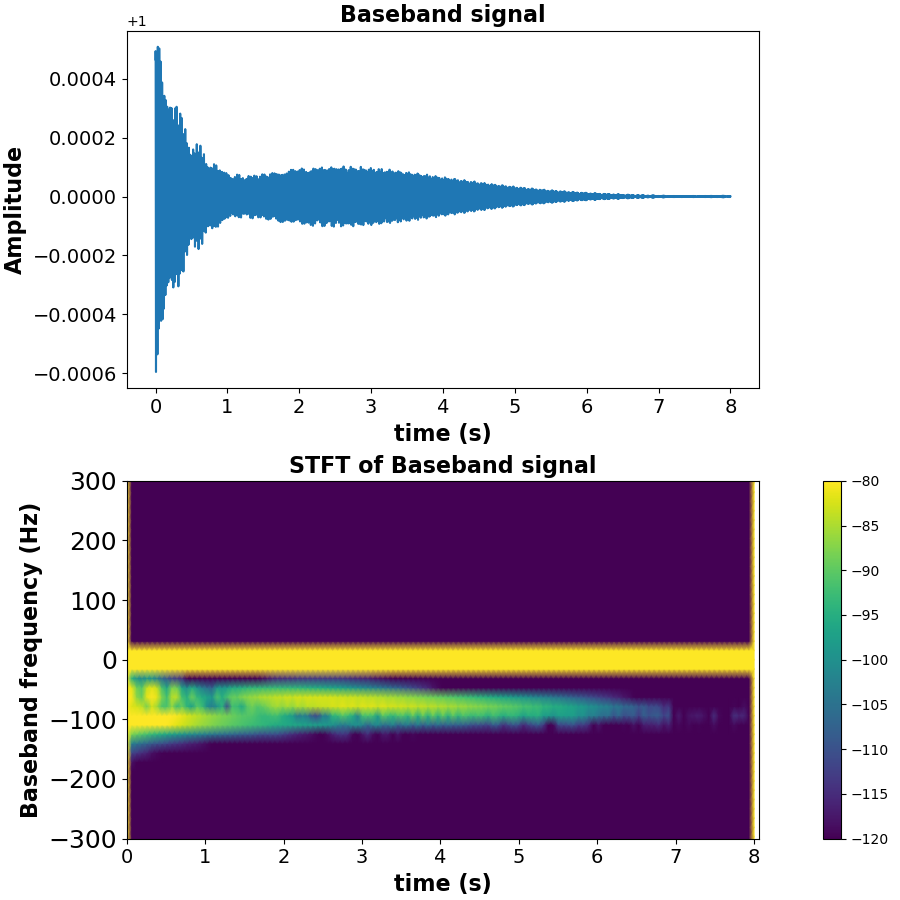}
		\caption{\label{fig5g}}
	\end{subfigure}

	\caption{(a) Motion Trajectory (b) In-plane directivity (c) Beam-plane directivity (c) Stepping velocity profile (e) Baseband signal and Doppler characteristics for stepping velocity profile (f) Constant velocity profile (g) Baseband signal and Doppler characteristics for constant velocity profile}
\end{figure}

In this section, we demonstrate the influence of wall reflections on the ultrasound signal characteristics. As mentioned earlier, we incorporate the reflections using the ISM method. To illustrate this, we consider the room environment with Wall 5 (z= 1.8 m) and Wall 6 (z=-1.2 m) as reflective with reflection coefficients of 0.8 each. We consider the piston source that is emitting upwards at an angle of 60 degrees from the x-y plane.   We depict the in-plane and the beam-plane directivities for such a source in Figures \ref{fig5b} and \ref{fig5c} respectively. Note that the polar angle is measured from the z-axis and hence the main-lobe of the  source directivity occurs at  $\phi =30^o$. The in-plane directivity values vary only by about 3-4 dB. We show results for two motion velocity profiles for walking on the axis away from the source as depicted in Figure \ref{fig5a}. These include motion with a stepping velocity profile and motion with a constant speed depicted in Figures \ref{fig5d} and \ref{fig5e} respectively. We first observe the characteristics of the baseband signal (Top panel) in Figures \ref{fig5e} and \ref{fig5f}, and compare them to those in Figures \ref{fig3d} and \ref{fig3f} respectively for the same motion trajectory. While the signals in \ref{fig3d} and \ref{fig3f} decay continuously, the signals in \ref{fig5d} and \ref{fig5e}  first decay and then increase as the target moves away from the source. This difference is due to presence of reflections from the top-wall (Wall-5) reaching the target contributing to the additional energy as the source in this case is directed upwards. The ray geometry of reflections dictate that the reflection from the Wall-5 reaches the x-axis at x=2.08 m. The target reaches this location at about 3.55 sec. Note that the reflection is composed of the main lobe with an angular spread and takes a longer-path resulting in further attenuation of the signal. The presence of strong reflection competes with the attenuation from the longer-path gives a peak in signal strength at about t=2.8 sec. Note that Figures \ref{fig5e} and \ref{fig5f} each show two traces for the Doppler characteristics in the STFT. The first one correponds to the direct-path from the source to the target and the second one corresponds to the image source from the wall reflection (Wall-5). This appears and gets stronger after about t=1.5 sec when the target reaches the location where the wall reflection is incident on the target. Also, the Doppler shift observed for the second trace is lower than that for the first trace. This is because the target is moving at an angle to the main lobe (beam) of the image source (imaged about Wall-5) but not on its axis resulting in a reduced Doppler shift compared to the trace for the direct-path signal. This example illlustrates the manifestation of reflections in the Doppler signal characteristics. In a reverberant  environment, it can be seen that multiple reflections result in Doppler characteristics to spread in frequency due to multiple image sources contributing to the signal. Moreover, the term ``Doppler frequency shift" as detemined just based on target velocity  loses relevance as it is not a single number but in fact a distribution dependent on the nature of reflections and the target motion in the environment.

\subsection{Rigid extended targets}
In this section, we present results from simulations on rigid extended targets i.e., targets of fintie spatial extent unlike the point targets discussed in the previous section. We model rigid extended targets as point clouds with each point defined by a scattering response function in terms of incident-direction ($\hat{n}_i$), scattered/reflected direction ($\hat{n}_r$) and the  target orientation ($\hat{n}_o$). This methodology allows one to incporprate the effects of multiple ray-paths from the source to the target and also the orientation-dependence of the scattered field from the target. The scattering response function could be measured or numerically simulated for a given target by using different plane-wave incident angles ($\hat{n}_i$) and estimating the scattered field in different reflected directions ($\hat{n}_r$). For extended targets, the orientation of the target plays a key role in the scattered response. To that end, in our simulations, we use the scattering response function defined by the following equations
\begin{eqnarray}
	\alpha (\hat{n}_i, \hat{n}_r; \hat{n}_o) &=& ||\alpha ((\hat{n}_i, \hat{n}_r; \hat{n}_o))|| e^{i\Phi(\hat{n}_i, \hat{n}_r; \hat{n}_o)} \\ \label{eqn45}
	||\alpha(\hat{n}_i, \hat{n}_r; \hat{n}_o)|| &=& (\alpha_{min}(1- (\hat{n}_i. \hat{n}_o)^2) + \alpha_{max} (\hat{n}_i. \hat{n}_o)^2) \left( \dfrac {e^{\left(\dfrac {(1-(\hat{n}_i. \hat{n}_r))} { 2 \sin^2(\theta_0)} \right)} -1 } {e^{\left(\dfrac {1} { \sin^2(\theta_0)} \right) }-1 } \right)\\\label{eqn46}
	\Phi(\hat{n}_i, \hat{n}_r; \hat{n}_o) &=& \omega_c \delta((1- (\hat{n}_i. \hat{n}_o)^2) + (1- (\hat{n}_r. \hat{n}_o)^2)) \label{eqn47}
\end{eqnarray}

The above choice incorporates the following aspects of physics in the scattering response:
\begin{enumerate}
	\item The scattered response has a preferred orientation and is the strongest when the incident direction is aligned (i.e., $\hat{n}_i = \pm \hat{n}_o$) with the target orientation and decays away from that preferred orientation. 
	\item For a given incident direction, the scattered response is strongest in the back scatter direction i..e., $\hat{n}_r = -\hat{n}_i$ 
	\item The scattering response has an angular spread about the back-scatter direction and decays away from it. The angle, $\theta_0$ in Eqn(\ref{eqn45}) dictates the angular spread of the scattered field around the backscatter direction.
	\item The phase function $\Phi(\hat{n}_i, \hat{n}_r; \hat{n}_o)$ accounts for the phase delays in the scattered response and is dependent on the target orientation, incident and reflected directions, and the carrier frequency.
\end{enumerate}
Note that the scattering response presented above is just one choice we use in this article and any physically meaningful or measured scattering response could be used with the simulation framework presented above.

Next, we present some simulation results with the above model for the rigid extended targets. For all the results presented in this subsection, we assume the source directivity to be uniform i.e. a spherical point source to delineate only the effect of target orientation in the simulated ultraosund signal characteristics. For the extended target, we consider a planar target with $1.2 \hspace{2pt} m$ height that is represented by 32 point discretization in the z-direction and orientation, $\hat{n}_o$ defined in the xy plane. Each point has a scattering function as defined above and represented with the parameters, $\alpha_{min} = 0.45$, $\alpha_{max} = 0.9$, $\theta_0$ =$\dfrac{\pi}{4}$, $\delta=0$, and $\omega_c = 2\pi*33000 \hspace {2pt} \mathrm{Hz}$.

\begin{figure}[H]
	\centering
	\begin{subfigure}[t]{0.4\textwidth}
		\centering
		\includegraphics[scale=0.15]{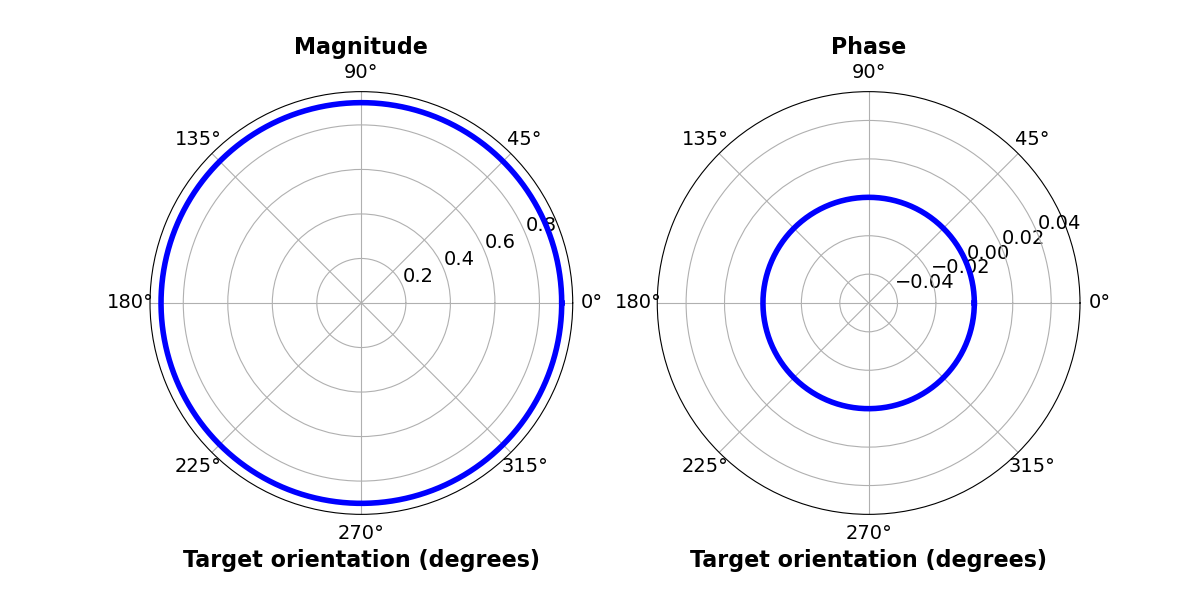}
		\caption{\label{fig6a}}
	\end{subfigure}
	\begin{subfigure}[t]{0.4\textwidth}
		\centering
		\includegraphics[scale=0.15]{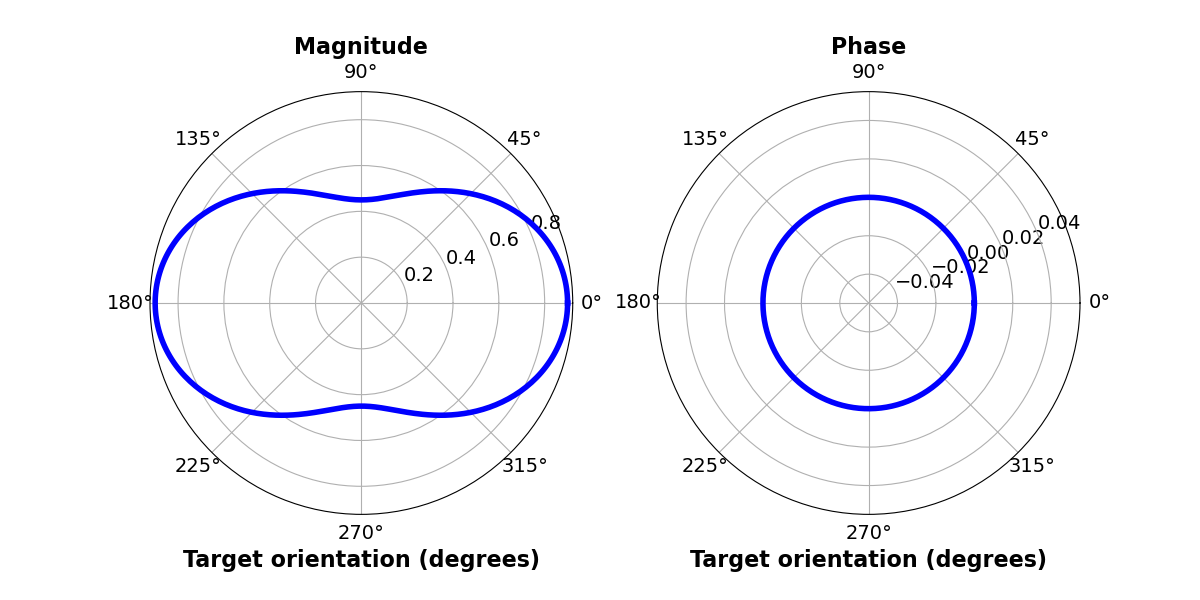}
		\caption{\label{fig6b}}
	\end{subfigure}
	\begin{subfigure}[t]{0.4\textwidth}
		\centering
		\includegraphics[scale=0.15]{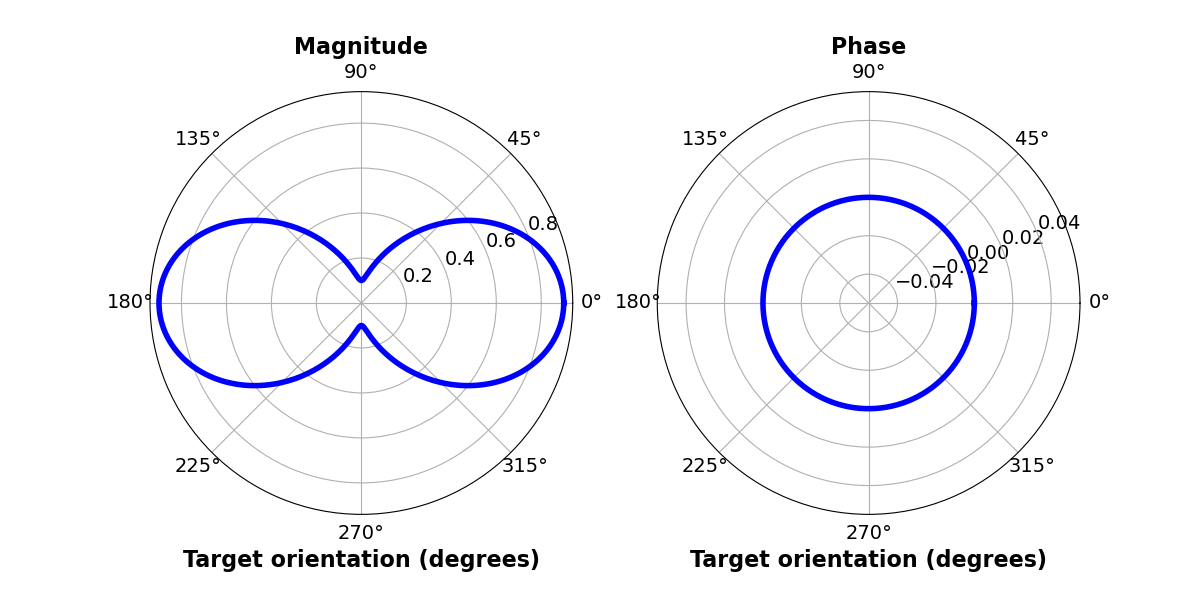}
		\caption{\label{fig6c}}
	\end{subfigure}
	
	\caption{Scattering response for $\hat{n}_i=(-1,0,0)$ and $\hat{n}_r=(1,0,0)$ for different target orientations, 0 to 360 degrees. (a) $\alpha_{min}=\alpha_{max}=0.9$ (b) $\alpha_{min}=0.45$, $\alpha_{max}=0.9$ (c) $\alpha_{min}=0.1, \alpha_{max}=0.9$ \label{fig6} }
\end{figure}

We first present the plots of the scattering response function used for each of the points. Figure \ref{fig6} shows the scattered response plots as a function of target orientation for different coefficients, $\alpha_{min}$ and $\alpha_{max}$. The incident and refelcted directions are fixed and given by  $\hat{n}_i=(-1,0,0)$ and $\hat{n}_r=(1,0,0)$. Notice that in Figure \ref{fig6a} when $\alpha_{min}=\alpha_{max}$ , we see that the scattered response is uniform irrespective of the target orientation. In Figures \ref{fig6b} \& \ref{fig6c}, as the difference between $\alpha_{min}$ and $\alpha_{max}$ increases the response becomes non-uniform with a maximum scattering response when $\hat{n}_i = \pm \hat{n}_o$. As mentioned earlier, these plots depict the backscatter response is the strongest when target is oriented to the incident direction and is minimum when it is oriented orthogonal to it.

\begin{figure}[H]
\centering
\includegraphics[scale=0.15]{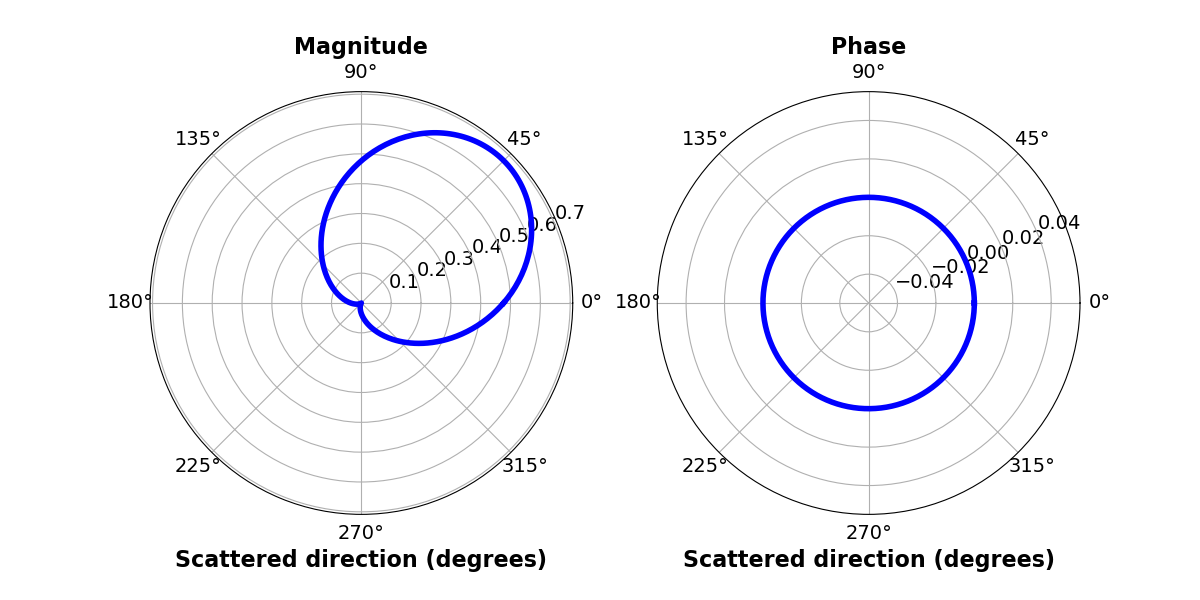}
\caption{Scattering response for incident direction, $\hat{n}_{i} = (-\frac{1}{\sqrt 2},-\frac{1}{\sqrt 2},0)$ and target-orientation, $\hat{n}_{o}=(1,0,0)$ for different scattered/reflected directions, 0 to 360 degrees. \label{fig7}}
\end{figure}

Figure \ref{fig7} shows the scattered response for an incident direction, $\hat{n}_{i} = (-\frac{1}{\sqrt 2},-\frac{1}{\sqrt 2},0)$ on a target oriented in the zero-degree direction i.e., $\hat{n}_{o}=(1,0,0)$. As can be seen, the scattered response is the strongest for $\hat{n}_r=-\hat{n}_i$ and decays away from that direction as described by the Eqn(\ref{eqn45}).  

\begin{figure}[H]
	\centering
	\begin{subfigure}[t]{0.4\textwidth}
		\centering
		\includegraphics[scale=0.15]{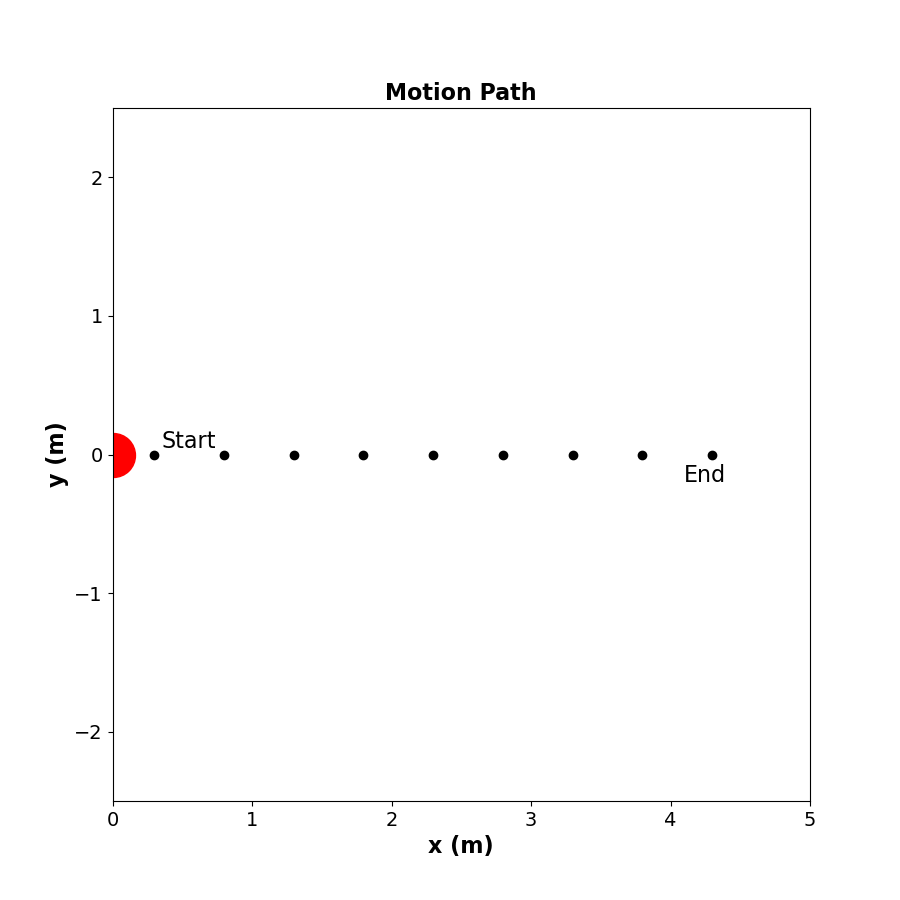}
		\caption{\label{fig8a}}
	\end{subfigure}
	\begin{subfigure}[t]{0.4\textwidth}
		\centering
		\includegraphics[scale=0.14]{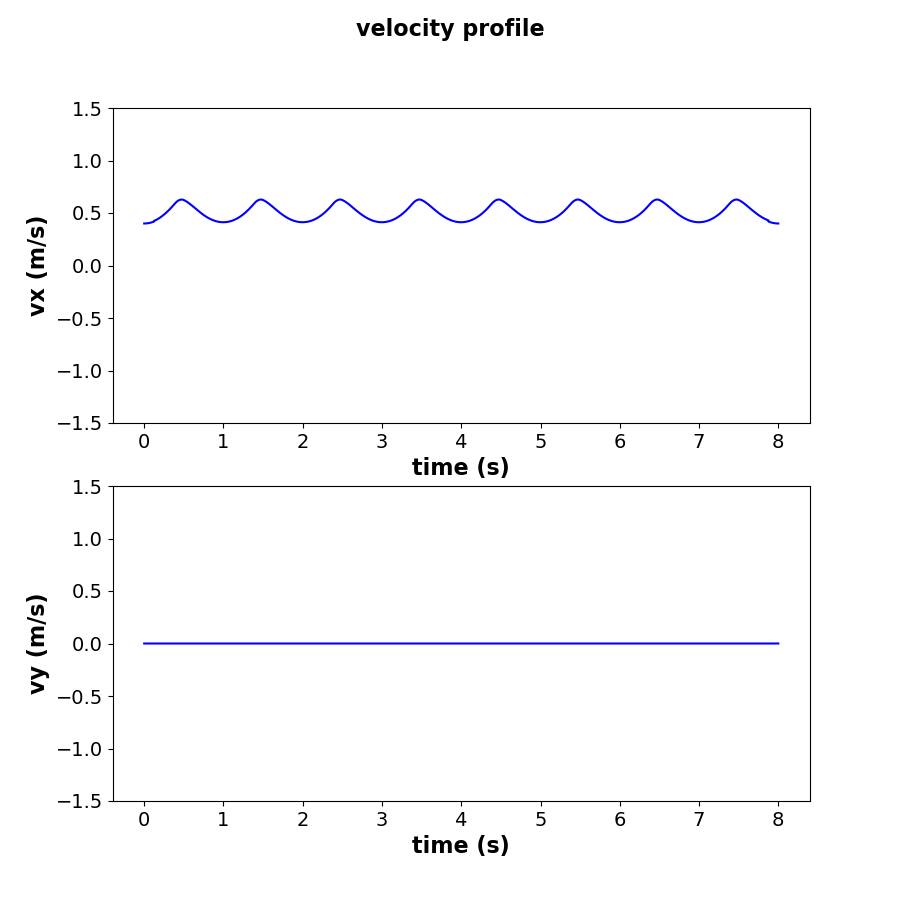}
		\caption{\label{fig8b}}
	\end{subfigure}
	\begin{subfigure}[t]{0.4\textwidth}
		\centering
		\includegraphics[scale=0.14]{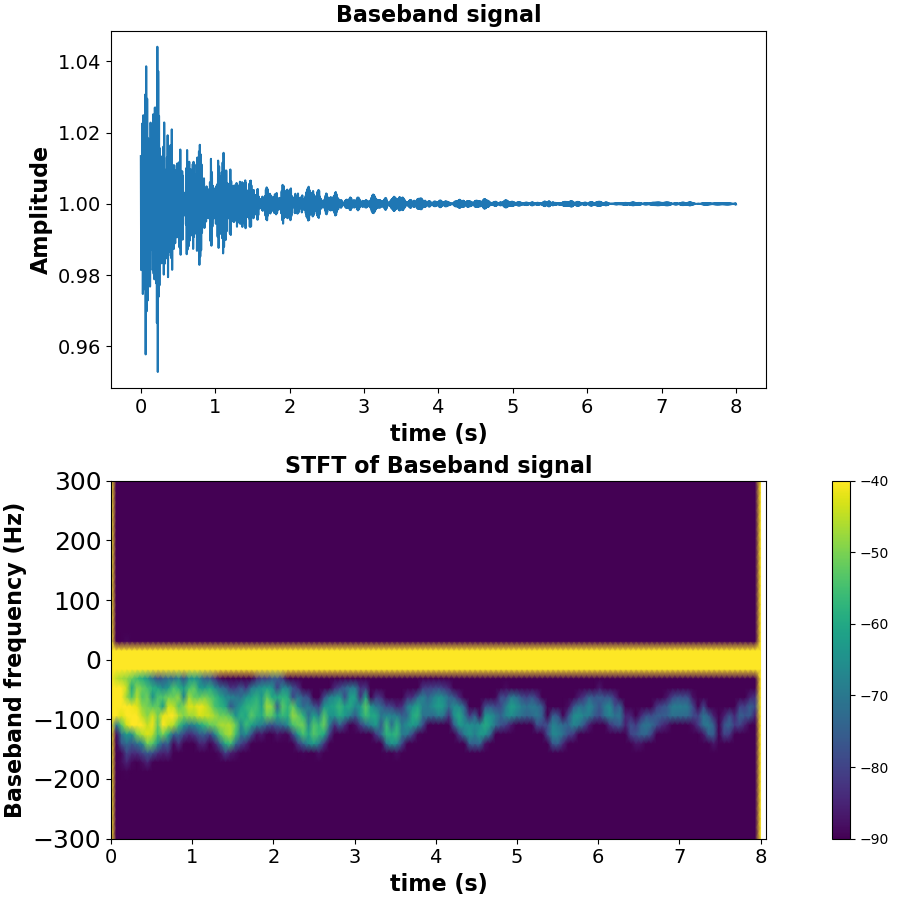}
		\caption{\label{fig8c}}
	\end{subfigure}
	
	\caption{(a) Motion trajectory b) Velocity Profile (d) Baseband doppler characteristics for stepping velocity profile.}
\end{figure}

First, we demonstrate the effect of having an extended target that is modeled as a point cloud. Below are the results for the case where the target is moving away form the source infront of it. We would like to compare this result with that from figure \ref{fig5d} where we can see a monotonic decay of the signal amplitude (Top -panel)  and clean  Doppler signal features pertaining to stepping velocity profiles. For the extended target case, while we see the decay in signal strength in Figure \ref{fig8c}, the Doppler signal features in Figure \ref{fig8c} are more smeared due to the presence of multiple points representing the extended target and scattering ultrasound. 

\begin{figure}[H]
	\centering
	\begin{subfigure}[t]{0.4\textwidth}
		\centering
		\includegraphics[scale=0.15]{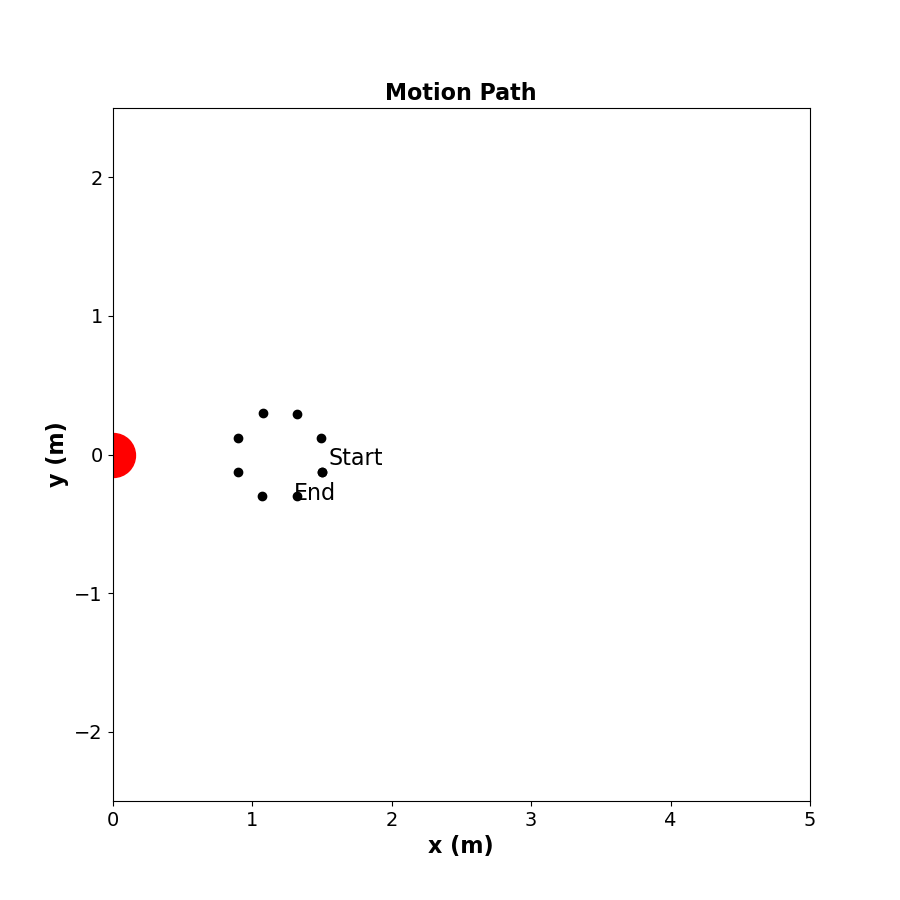}
		\caption{\label{fig9a}}
	\end{subfigure}
	\begin{subfigure}[t]{0.4\textwidth}
		\centering
		\includegraphics[scale=0.14]{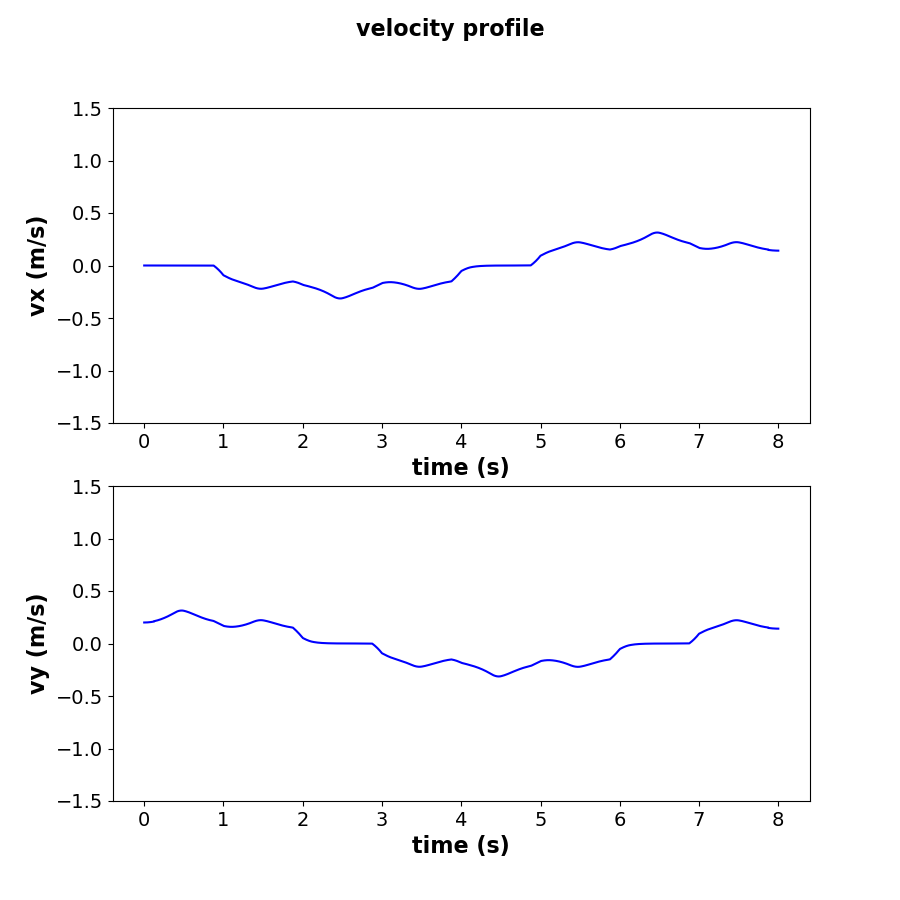}
		\caption{\label{fig9b}}
	\end{subfigure}
		\begin{subfigure}[t]{0.4\textwidth}
		\centering
		\includegraphics[scale=0.14]{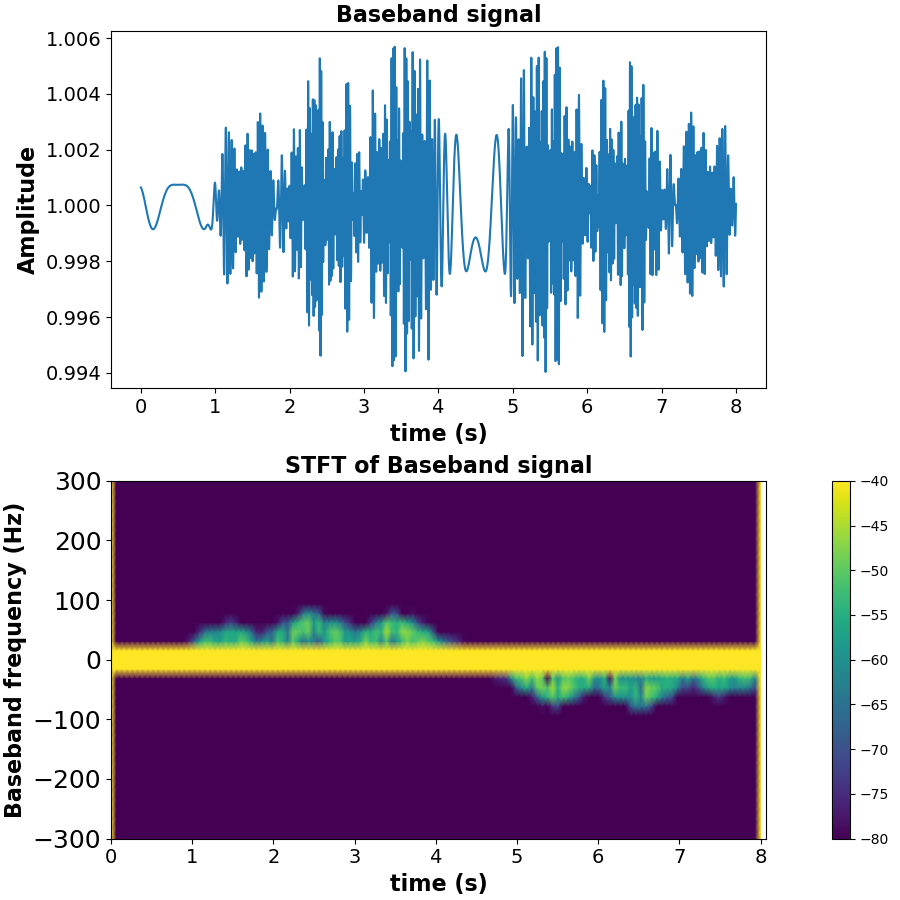}
		\caption{\label{fig9c}}
	\end{subfigure}

	\caption{(a) Circular motion trajectory b) Velocity Profile (d) Baseband signal and Doppler characteristics for circular motion trajectory.}
\end{figure}

Next, we show an example to elucidate the impact of orientation-dependent scattered response on Doppler signals. We consider the case where a target is turning around in a circle on the axis of the source at around  $1.2 \hspace{2pt} m$ from it. The motion is reperesented by the target moving in a small circle of $0.3 \hspace{2pt} m$ radius centered at $1.2 \hspace{2pt} m$. Note that as the target is almost at a fixed distance from the source, variations in the signal strength are due to the chaging orientation of the target with respect to the source/recevier. We model the circular motion trajectory of the target as an octagon with piecewise linear segments. As can be seen in figure \ref{fig9c}, the baseband signal starts with zero Doppler shift (target moving on the octagon edge perpendicular to the source) and goes through different values during the course of the motion. Note that the amplitude of the signal increases and is higher when the user is moving in non-orthogonal direction to the source than when it is moving perpendicular to the source. In fact, we see four distinct amplitude levels pertaining to the edges of the octagon used to model the circular motion due to the change in the orientation of the target during motion.

\section{Experimental results: Qualitative comparison to the simulated baseband and Doppler signal characteristics \label{sec7}}
\begin{figure}[H]
	\centering
	\includegraphics[scale=0.12]{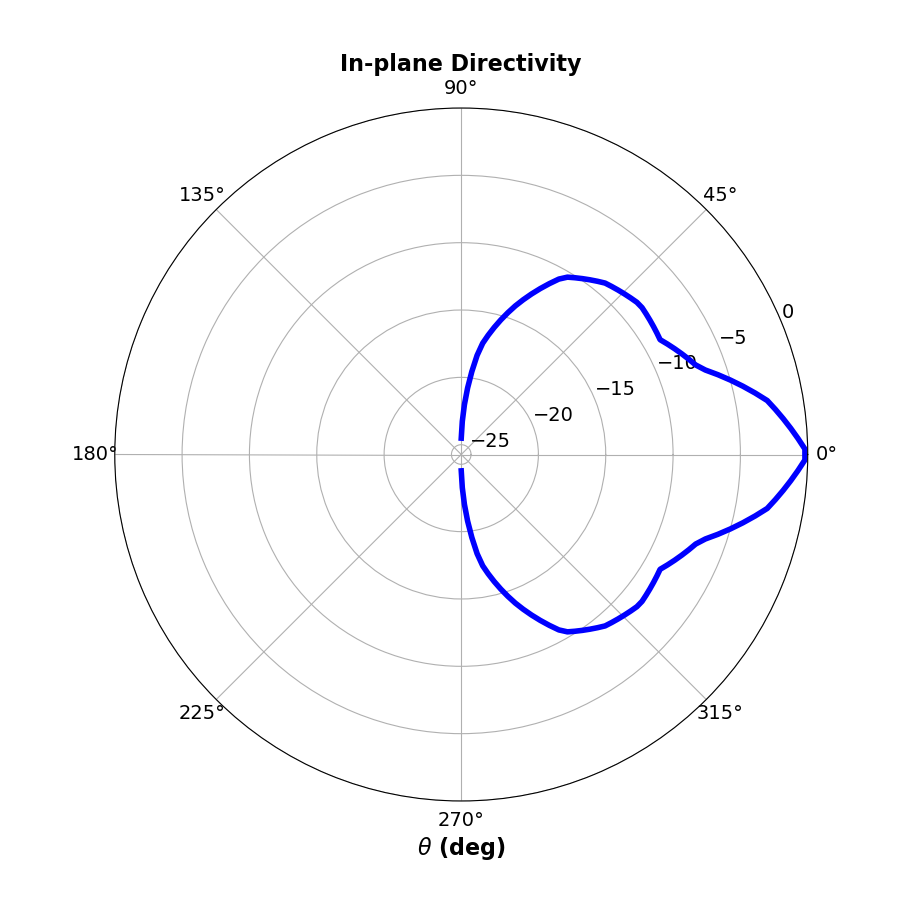}
	\caption{Measured in-plane directivity of the source at 33 kHz. \label{fig10}}
\end{figure}

\begin{figure}[H]
	\centering
	\begin{subfigure}[t]{0.33\textwidth}
		\centering
		\includegraphics[scale=0.16]{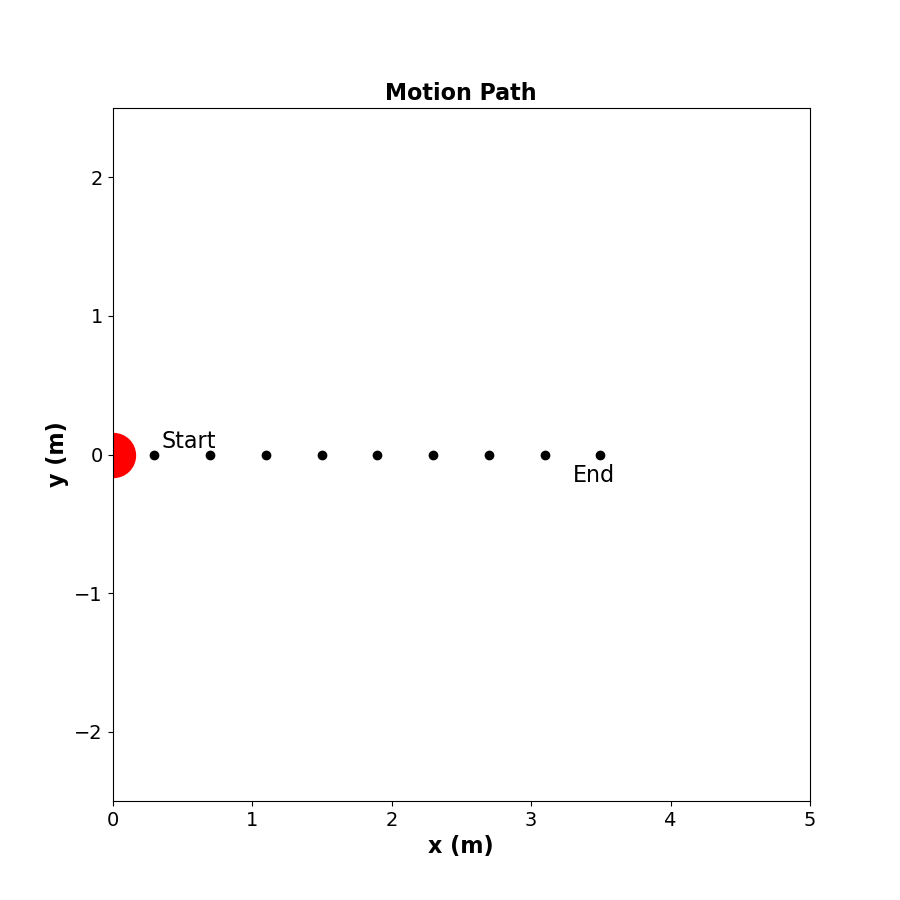}
		\caption{\label{fig11a}}
	\end{subfigure}
	\begin{subfigure}[t]{0.33\textwidth}
		\centering
		\includegraphics[scale=0.21]{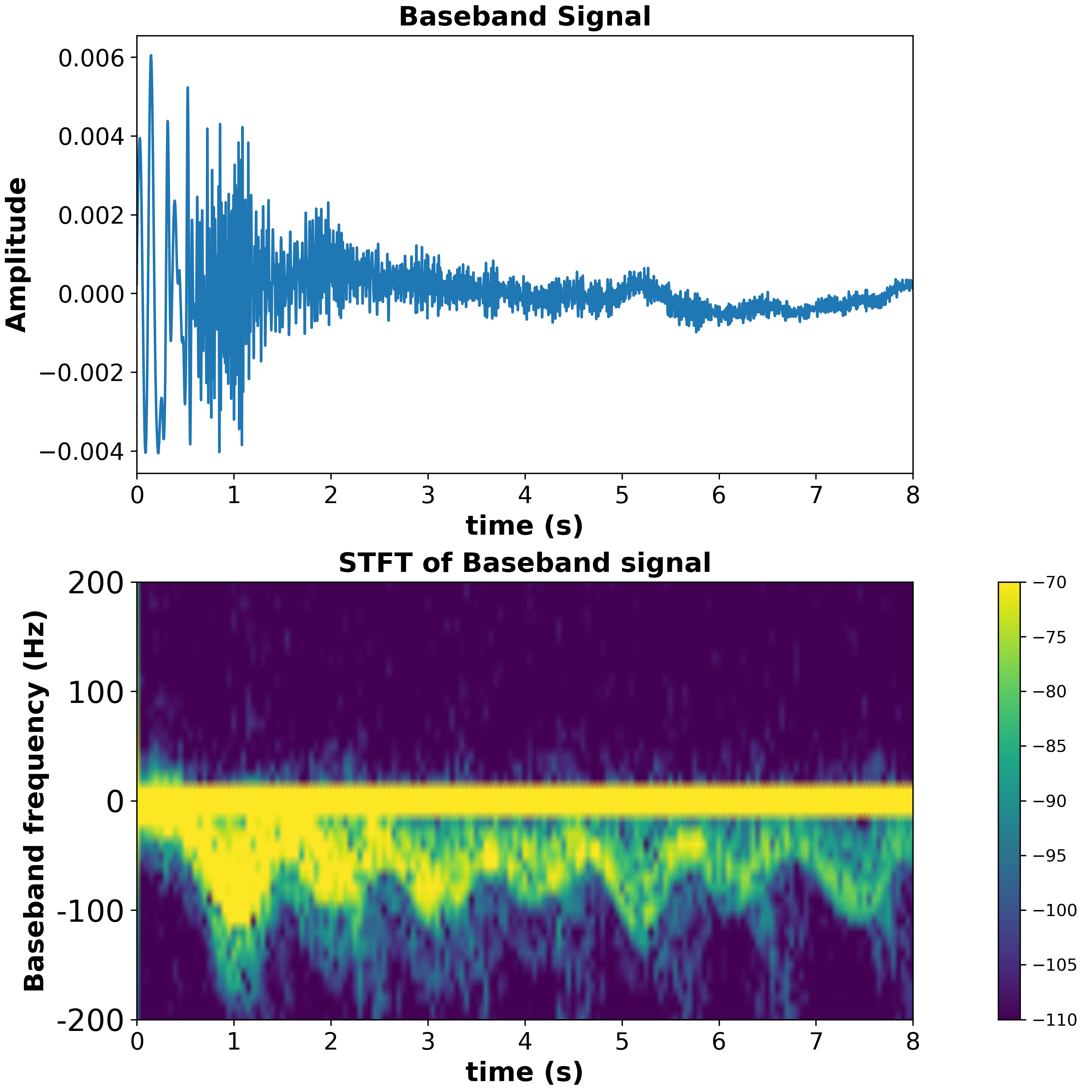}
		\caption{\label{fig11b}}
	\end{subfigure}
	\begin{subfigure}[t]{0.33\textwidth}
		\centering
		\includegraphics[scale=0.15]{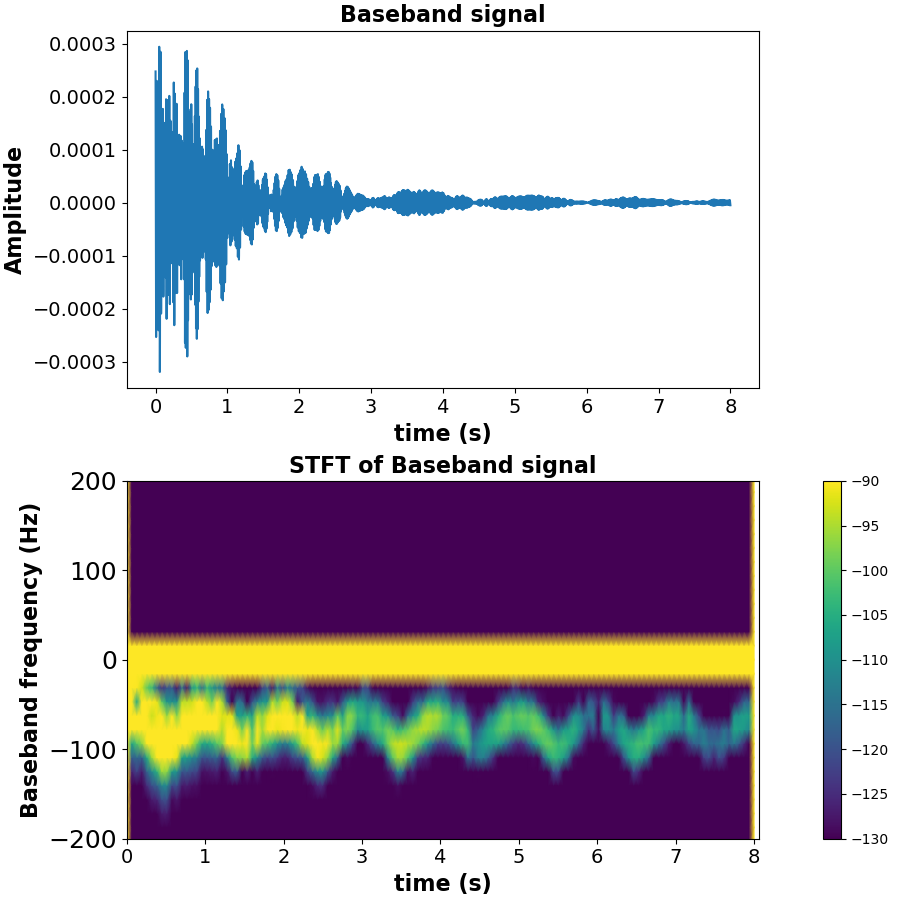}
		\caption{\label{fig11c}}
	\end{subfigure}
	
	\begin{subfigure}[t]{0.33\textwidth}
		\centering
		\includegraphics[scale=0.16]{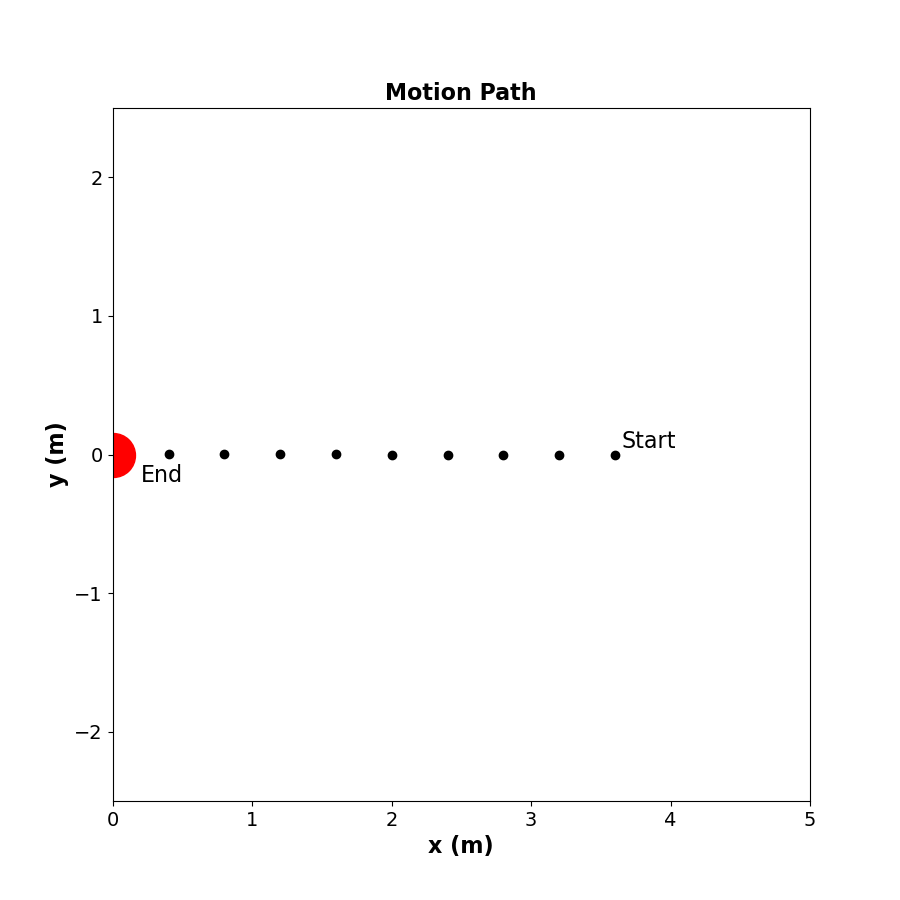}
		\caption{\label{fig11d}}
	\end{subfigure}
	\begin{subfigure}[t]{0.33\textwidth}
		\centering
		\includegraphics[scale=0.21]{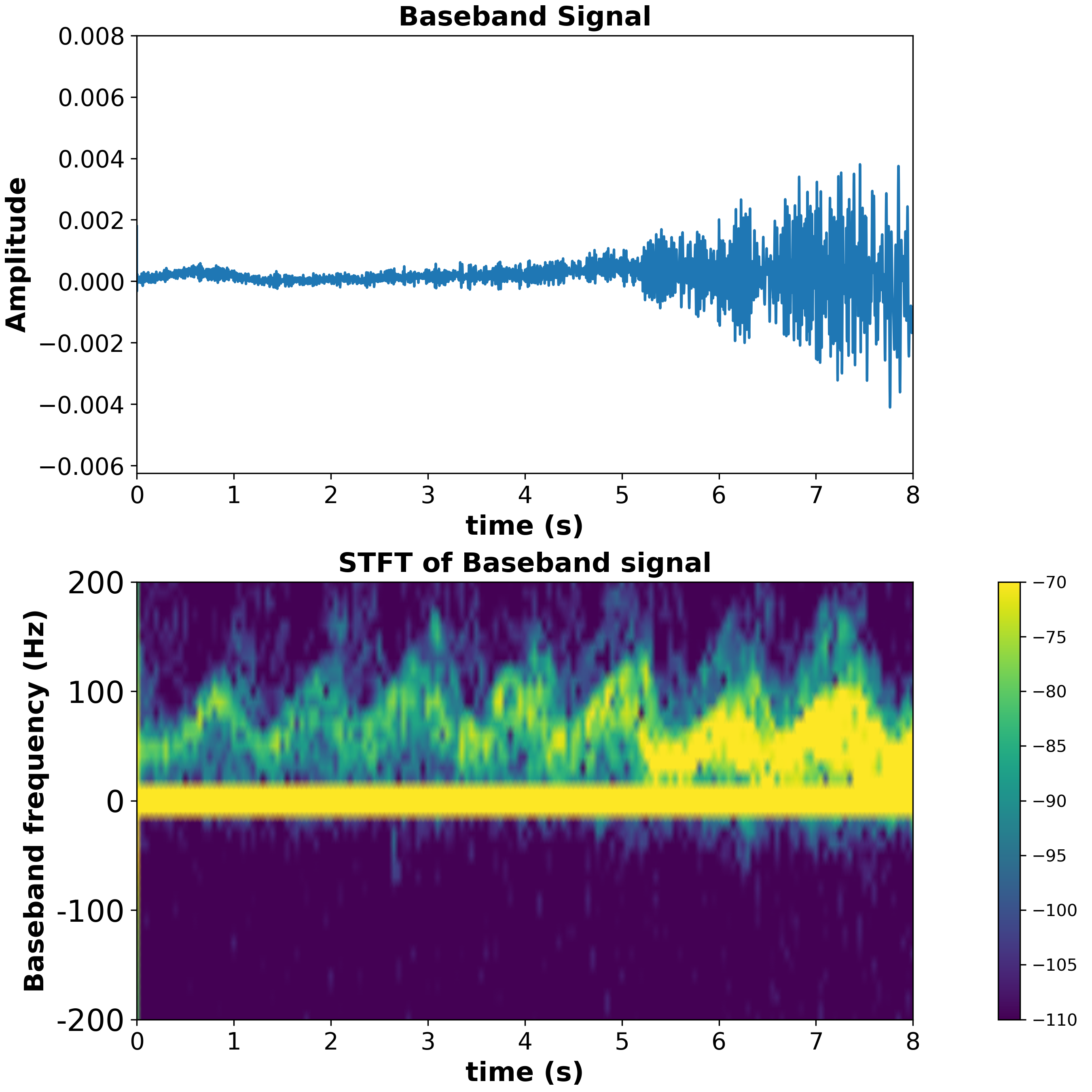}
		\caption{\label{fig11e}}
	\end{subfigure}
	\begin{subfigure}[t]{0.33\textwidth}
		\centering
		\includegraphics[scale=0.15]{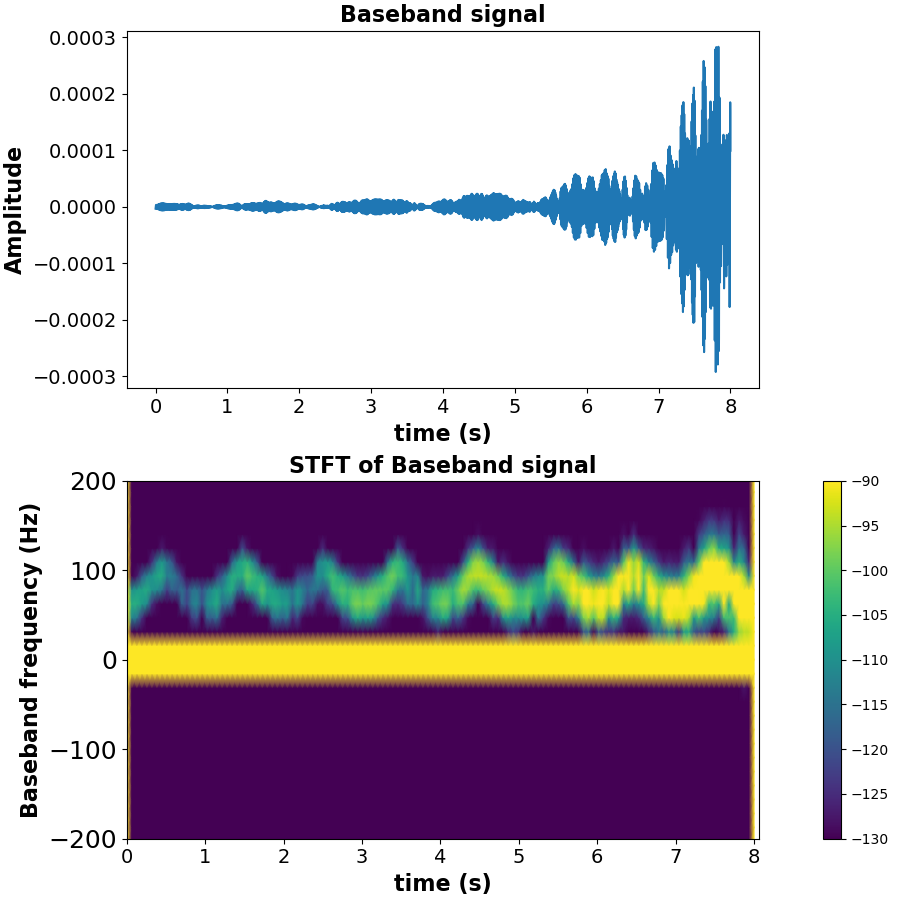}
		\caption{\label{fig11f}}
	\end{subfigure}
	
	\begin{subfigure}[t]{0.33\textwidth}
		\centering
		\includegraphics[scale=0.16]{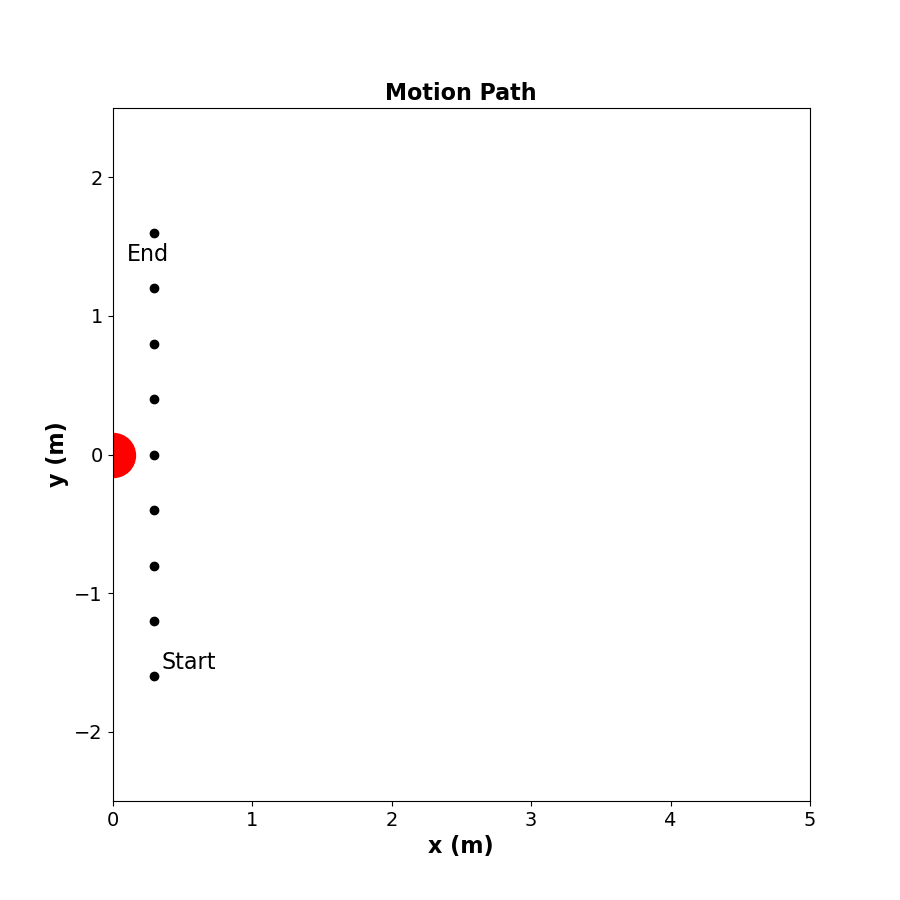}
		\caption{\label{fig11g}}
	\end{subfigure}
	\begin{subfigure}[t]{0.33\textwidth}
		\centering
		\includegraphics[scale=0.21]{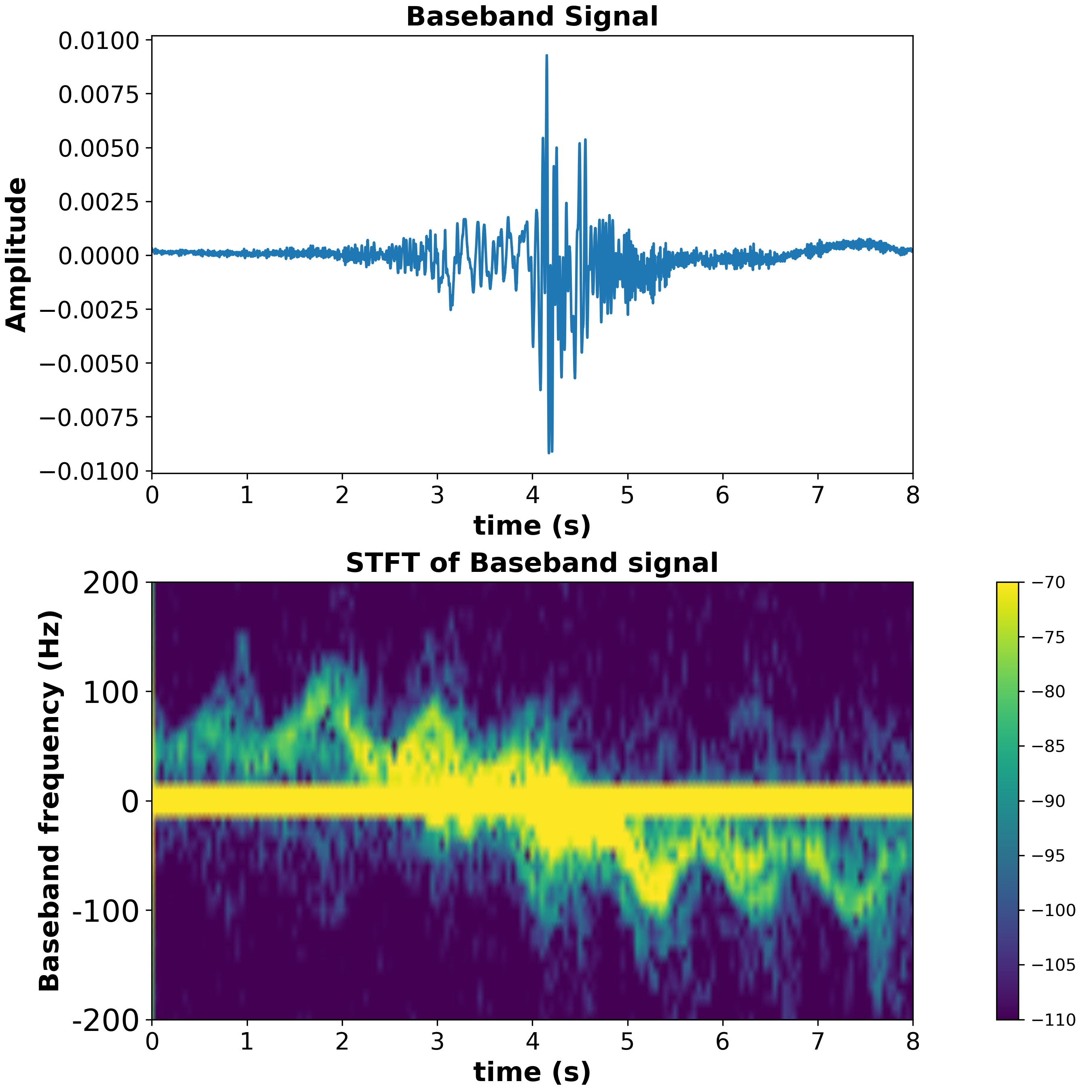}
		\caption{\label{fig11h}}
	\end{subfigure}
	\begin{subfigure}[t]{0.33\textwidth}
	\centering
	\includegraphics[scale=0.15]{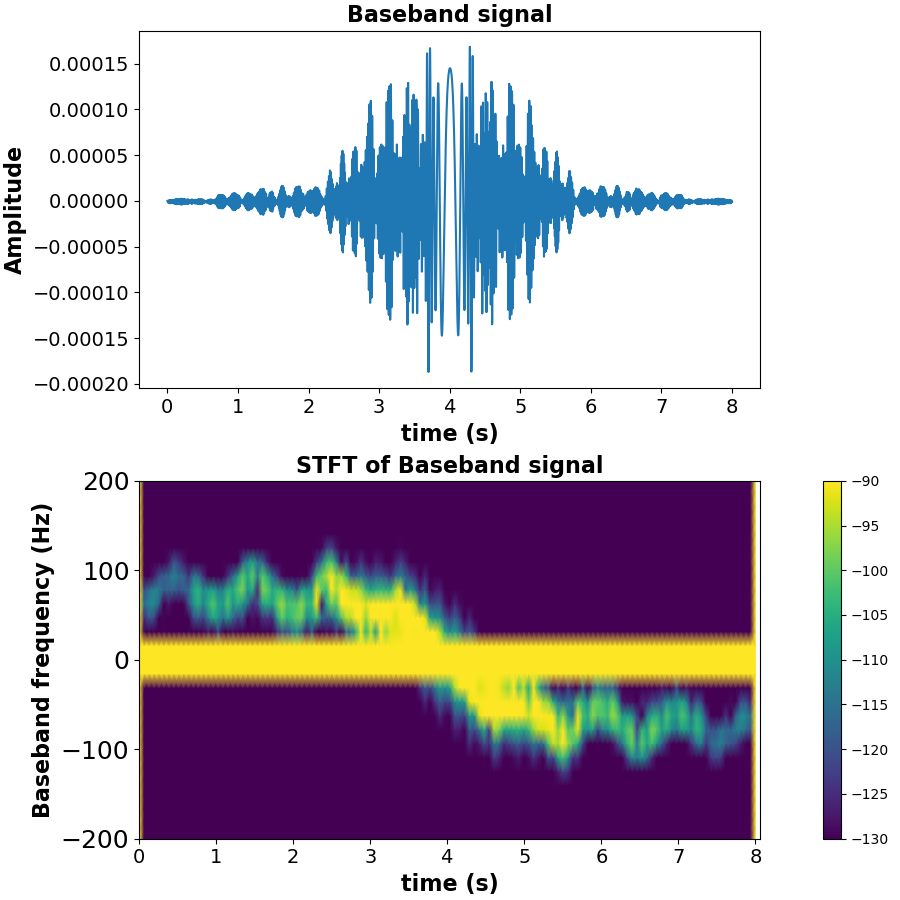}
	\caption{\label{fig11i}}
   \end{subfigure}
	
	\begin{subfigure}[t]{0.33\textwidth}
		\centering
		\includegraphics[scale=0.16]{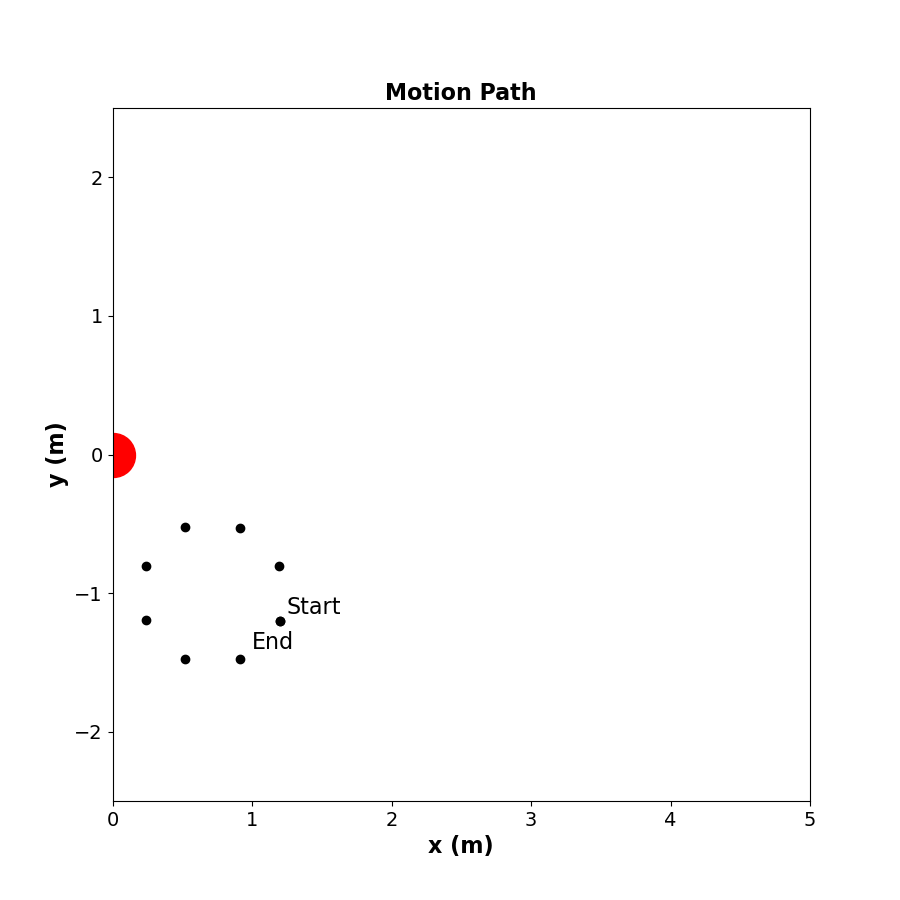}
		\caption{\label{fig11j}}
	\end{subfigure}
	\begin{subfigure}[t]{0.33\textwidth}
		\centering
		\includegraphics[scale=0.21]{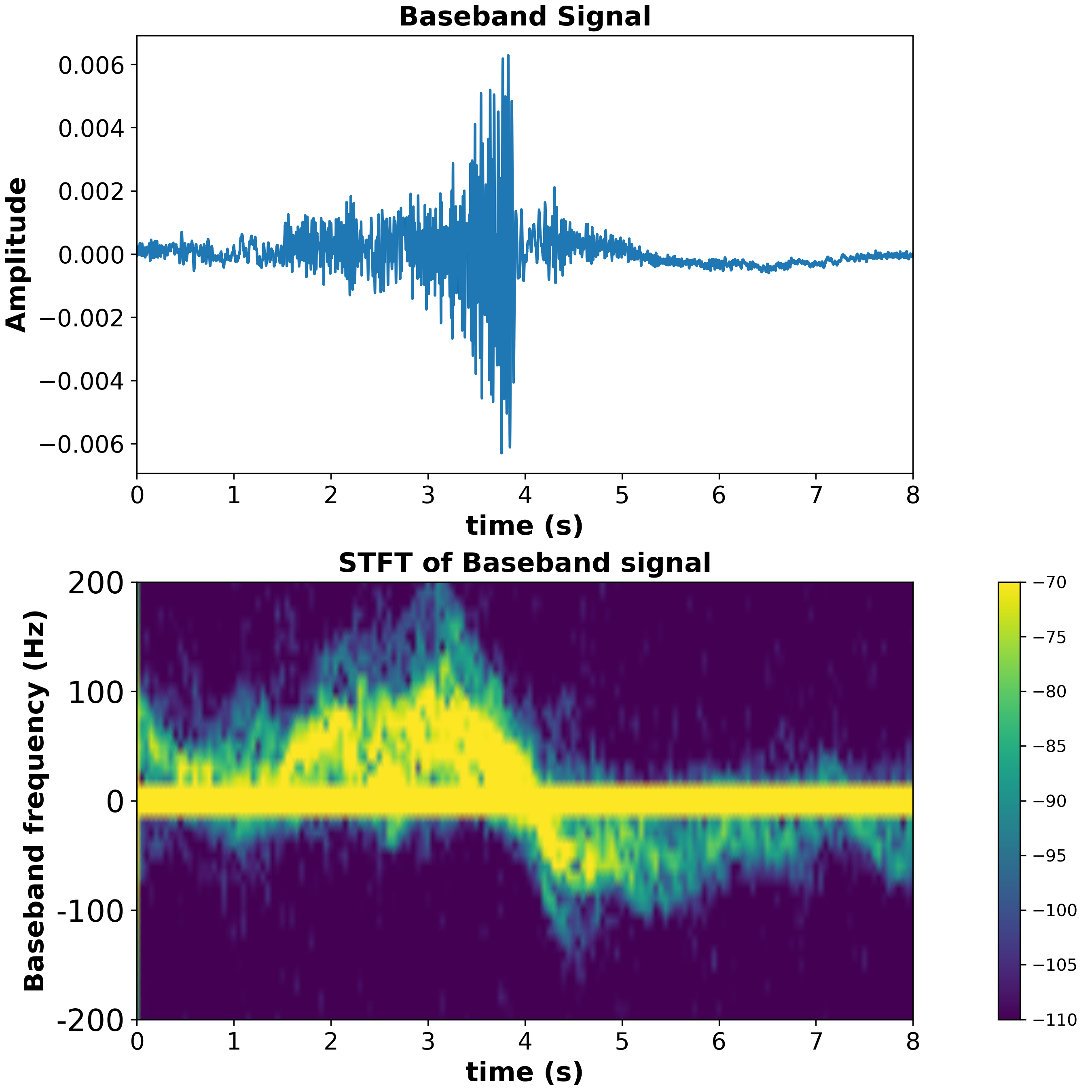}
		\caption{\label{fig11k}}
	\end{subfigure}
		\begin{subfigure}[t]{0.33\textwidth}
		\centering
		\includegraphics[scale=0.15]{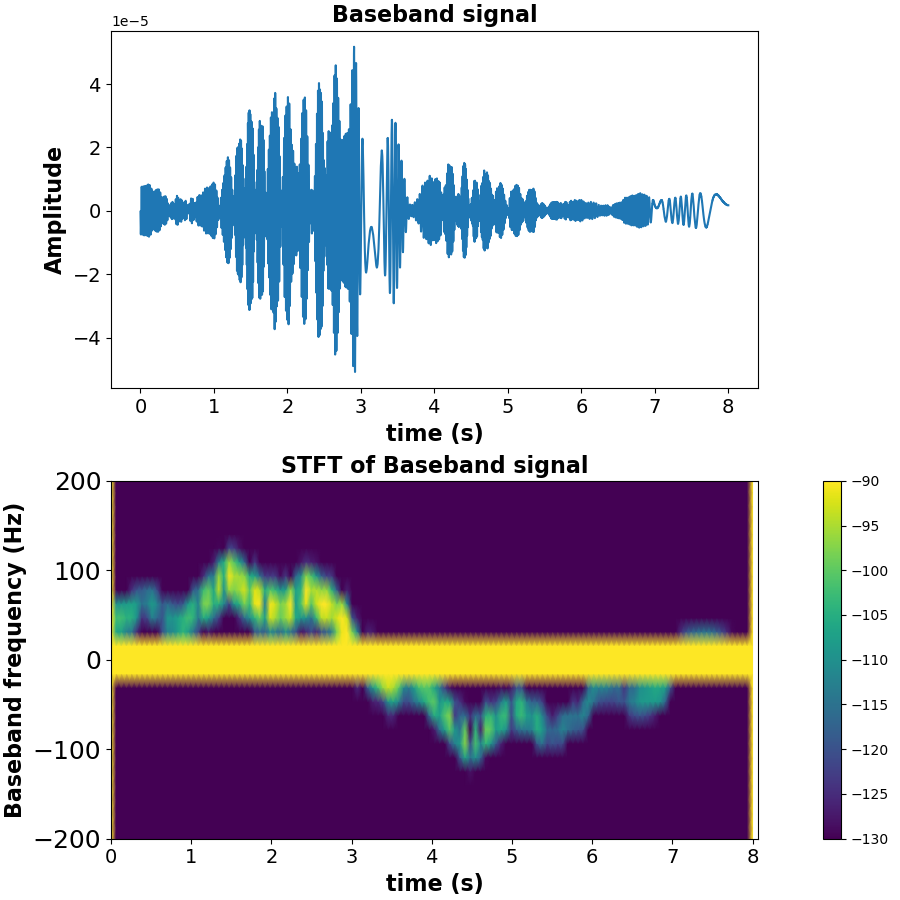}
		\caption{\label{fig11l}}
	\end{subfigure}
	
	\caption{Walking on-axis away---(a) Motion trajectory (b) Experiment (c) Simulation; Walking on-axis towards---(d) Motion trajectory (e) Experiment (f) Simulation ; Walking transverse at 0.3 m -- (g) Motion trajectory (h) Experiment (i) Simulation; Walking in circle ---(j) Motion trajectory (k) Experiment (l) Simulation \label{fig11}} 
\end{figure}

In this section, we present  experimental results for a few representative motion scenarios and compare the ultrasound signal characteristics observed in the experiments with those obtained from the simulations using the methodology presented in the article. The experiments were conducted using a high-frequency acoustic source (radius=7 mm) that has directivity as shown in the Figure \ref{fig10} at 33 kHz measured with a BRÜEL \& KAER microphone (B\&K-4939) at 10 cm from the source in the horizontal plane. 
For the experimental data with target motion, the microphone is positioned about 5 cm above the source and the microphone signals were captured at a sampling rate of 96 kHz. The target is an actual human moving in prescribed motion trajectories in a room-environment with the source and receiver/microphone. The room used for experimental data collection has the dimenstions $8 m \hspace {2 pt} \mathrm{(length)} \times 5 m \hspace {2 pt} \mathrm{(width)} \times 2.5 m \hspace {2 pt} \mathrm{(height)}$. For the simulation, we use an extended target model with parameters described in the previous section. Note that we are offering only a qualitative comparison of the Doppler signal characteristics between simulations and experiments for the following reasons:
\begin{enumerate}
	\item The room-acoustic environment cannot be accurately replicated in the simulation and is especially challenging at the ultrasonic frequencies of interest for motion sensing.
	\item The high-frequency near-field effects from the source to the receiver/microphone positioned close to the source is not modeled accurately by the geometrical acoustics approximation. This results in a discrepancy on the strength of the direct path signal between simulation and experiment especially when the target is in the near field. 
	\item  The experimental data is acquired with a real human walking in the environment, the motion characteristics for which cannot be accurately replicated. While we are working on incoporating realistic human motions into our simulation framework, it imposes additional challenges from a simulation standpoint that we hope to discuss as part of future work.
\end{enumerate}
The motion trajectories from the experimental data collection are replicated in simulation and we compare the results for the baseband signal and the associated Doppler chartacteristics pertaining to the motion. For all the results shown here, we assume a stepping velocity profile ($v_{min}=0.32 \hspace{2pt} m/s, v_{max}=0.56 \hspace{2pt} m/s$) with an average speed of $v_{avg}=0.4 \hspace{2pt} m/s$ that corresponds well with the nominal walking speed used for the experimental data collection. Figure \ref{fig11} shows the comparison results for four different walking scenarios: 
\begin{enumerate}
	\item \underline{Walking away from the source}: Figure \ref{fig11a} shows the motion trajectory used for data collection and simulation. Figure \ref{fig11b} shows the experimental results and Figure \ref{fig11c} shows the simulated results, namely the baseband signal and the corresponding STFT showing the Doppler frequency characteristics observed for the motion trajectory. As can be seen, both the simulated and experimental results show the baseband signal (top panel) decaying away from the source. In addition, the STFT shown in bottom panel of Figure \ref{fig11c} for simulation reasonably captures the Doppler characteristics pertaining to walking observed in the experiments (Figure \ref{fig11b}). Note that the STFT's for both simulation and experiment are shown with a color-scale spanning 40 dB range to facilitate comparison. The experimental results in Figure \ref{fig11b} shows smearing of the Doppler frequency characteristics due to the fact that realistic human walking is not a rigid body motion and different parts of the body are in relative motion with one another unlike the walking profile used in the simulations which corresponds to a rigid body motion. 
	
	\item \underline{Walking towards the source}: Figures (\ref{fig11d}-\ref{fig11f}) shows the motion trajectory and the results obtained for the case where the target is approaching the source from about $3.6 \hspace{2pt} m$. This is the opposite scenario to that discussed above and as can be seen the signal strength increases in both simulation and experiment and the STFT from simulation in Figure \ref{fig11f} reasonably replicates the Doppler signal characteristics observed in the experiment (Figure \ref{fig11e}). Also, observe that the numerical value (including the sign) of the Doppler shifts observed for experiments agree well with that obtained in the simulations. 
	
	\item \underline{Walking transverse to the source at 0.3 m from the source}: In this example, we consider the case where the target is moving transverse to the source first approaching the source and then receding away from it as shown in Figure \ref{fig11g}. As can be seen in Figure \ref{fig11h}, the Doppler shift is postive for the first-half of the motion trajectory and then it switches to negative as the target moves away from the source.   Accrodingly, the signal strength increases and decreases as the target moves through the point of zero-Doppler shift. In other words, the baseband signal is the strongest closer to the point where the Doppler shift is zero. This is well captured in the simulation results depicted in Figure \ref{fig11i} where the signal strength increases reaching a peak and then decreases as the target moves away. Moreover, the range of Doppler shifts observed in the baseband signal for experiments agrees well with those observed in the simulation.
	
	\item \underline{Walking in a circle on one side of the source}:  This example compares the results for the case where the target is walking in a circle on one side of the source as depicted in Figure \ref{fig11j}. In the experiment, the target starts and moves in a circle towards the source and then recedes away from it. Hence, the Doppler shifts observed in the Figure \ref{fig11k} are positive for the first half of the motion and then negative for the next half of the motion. Also, notice that the signal strength increases during the first half and then reduces for the next half of the motion duration. Note that the near-field effects are predominant for this case as the target is moving close to the source. We see similar observations from the simulated results where the patterns observed for Doppler shifts in Figure \ref{fig11l} agree well with those from the experiment. As noted above, the near field effects coupled with non-rigid body type motions in the experiments result in broader smearing or distribution of the Doppler frequencies that are not well-captured in the simulation. 
\end{enumerate}

 The above results and comparisons demonstrate the feasibility of employing the proposed methodology to simulate the physics of ultrasound based motion sensing that enables one to generate synthetic data that captures key aspects signal characteristics pertinent to the Doppler based motion sensing. Incorporating more realisitc human motions and enhanced simulation capabilities for airborne ultrasound in room-acoustic environments will enhance the accuracy of the proposed methodology and is a work underway. 

\section{Conclusions \label{sec8}}

In this article, we presented a numerical simulation methodology for ultrasound Doppler based motion sensing applications. The approach relies on an analytical formulation developed in Sections (\ref{sec2}-\ref{sec4}) where the motion sensing process was represented in two parts: i) moving point target first acts as a receiver and expressions are obtained for the incident field on the target and then ii)  the target is treated as a moving source scattering the incident field back to the receiver/microphone and the corresponding expressions for the received signal are derived. The analytic-signal representation of the received signal was used to obtain explicit expressions for the instantaneous signal amplitude and phase that in general correspond to Amplitude (AM) and Frequency Modulation (FM) observed in the signals.  In particular, the expressions for Doppler shift were derived from the instantaneous frequency computed as the time-derivative of the inastantaneous phase and is shown to match the well-known expressions for Doppler shift for a few simple cases. In section \ref{sec5}, the approach was extended to incoporate non-uniform source directivity and multiple reflections using the Image Source Method widely used in Room Acoustics. Since ultrasonic energy decays faster due to higher attenuation compared to the audible frequencies in indoor environments, first order image source method provides a reasonably good approximation for ultrasound based motion sensing applications. Next, we discussed the numerical implementation aspects of the proposed methodology highlighting how the baseband signal implementation allows one to run the simulations at much coarser time-deiscretization, thereby allowing one to implement the methodology for longer motion durations and different motion trajectories. In section \ref{sec6}, we presented results with the motion of a point target in a room highlighting the influence of source directivities, velocity profiles, and motion trajectories on the Doppler signal characteristics observed in the basedband signal. We then discussed an approach to represent extended targets as point-clouds with scattering response functions parametrized by the target orientation. Inclusion of the multiple points in the point clouds incorporates multiple ray-paths from the source to the target and contributes to modeling the smearing and spread in Doppler frequency spectrum that is typically observed in the experiements. In section \ref{sec7}, we demonstrated the capability of the simulation methodology to capture realistic motion signatures by offering a qualitative comparison between experimental data and the simulated outputs using the proposed approach. We demonstrated that the proposed approach captures key features pertaining to signal strength and Doppler shift characteristics for different motion scenarios. We note that there is no readily available methodology to simulate airborne ultrasound based motion sensing and the proposed technique could potentially pave the way to develop and improve such methodologies. In that respect, we would like to note the following advantages of the proposed simulation approach:
\begin{enumerate}
	\item The approach allows us to incorporate complex sources with different source directivities and wall reflection characteristics in a room environment through the Image Source Method.
	\item The approach is very accurate in terms of capturing motion related signatures for different motion scenarios and hence is very useful for generating large-scale feature-rich synthetic data for training machine learning algorithms for motion sensing applications. 
	\item Baseband implementation proposed here allows us to run simulations at significantly reduced computational costs without compromising the accuracy.
\end{enumerate}

Next, we would like to highlight some of the limitations of the proposed approach:
\begin{itemize}
	\item Due to the Geometrical Acoustics assumption, the  near-field effects when the target is close to the source/receiver are not accurately captured in the current simulation framework. 
	\item While we model an extended target with point-clouds and scattering response functions, the points are still constrained to move as a rigid body. This isnt realistic enough to replicate real motion scenarios.  We are actively working towards improving this to incorporate realistic motions by relying on tools for accurate motion rendering and further integrating this simulator with such tools.
\end{itemize}
The research and implementation work needed to address the limitations listed above are currently underway. We would like to end this article on the note that this is a first-step towards developing an integrated simulator for ultrasound motion sensing that is general enough to incorporate different sources, reflection environments, and motion trajectories that can be run at low computational costs while delivering reasonably accurate motion signatures as observed by ultrasound.

{\tiny }
\bibliographystyle{elsarticle-num}
\bibliography{bib_motion_sensing.bib}

\end{document}